\documentclass[aps,prx,twocolumn,nofootinbib,longbibliography,notitlepage,floatfix]{revtex4-2}
\usepackage{graphicx}
\usepackage[section]{placeins}
\usepackage{tabularx}
\usepackage{amsmath}
\usepackage{amstext}
\usepackage{amssymb}
\usepackage{xfrac}
\usepackage[colorlinks,citecolor=blue]{hyperref}
\usepackage{graphicx}
\usepackage{amsmath}
\usepackage{amstext}
\usepackage{amssymb}
\usepackage{amsfonts}
\usepackage{longtable,booktabs}
\usepackage{hyperref}
\usepackage{url}
\usepackage{subfigure}
\usepackage{dsfont}
\usepackage{booktabs}
\usepackage{amsbsy}
\usepackage{dcolumn}
\usepackage{amsthm}
\usepackage{bm}
\usepackage{esint}
\usepackage{multirow}
\usepackage{hyperref}
\usepackage{cleveref}
\usepackage{mathrsfs}
\usepackage{amsfonts}
\usepackage{amsbsy}
\usepackage{dcolumn}
\usepackage{bm}
\usepackage{multirow}
\usepackage{color}
\usepackage{extarrows}
\usepackage{datetime}
\usepackage{comment}

\usepackage{braket}
\usepackage{indentfirst}
\usepackage{cases}
\usepackage{diagbox}
\usepackage{multirow}
\usepackage{makecell}
\usepackage{blkarray}
\usepackage{longtable}
\usepackage{extarrows}

\usepackage{relsize}
\usepackage[utf8]{inputenc}   
\usepackage[T1]{fontenc}      

 \usepackage{datetime}
\usepackage{comment}
\usepackage[super]{nth}
\usepackage{tikz}
\usepackage{makecell}
\usepackage{aligned-overset}
\usetikzlibrary{arrows.meta}
\usetikzlibrary{positioning}
\tikzset{
    wcirc/.style={draw, circle, minimum size=1ex, inner sep=0pt},
    bcirc/.style={fill, circle, minimum size=1ex, inner sep=0pt},
	wcircL/.style={draw, circle, minimum size=1.5ex, inner sep=0pt},
    bcircL/.style={fill, circle, minimum size=1.5ex, inner sep=0pt}
}
\hypersetup{
	colorlinks=blue,
	linkcolor=blue,
	filecolor=blue,
	urlcolor=blue,
}
\usepackage{scalerel}
\usepackage{tikz}
\usetikzlibrary{svg.path}

\def\i{\mathrm{i}}

\definecolor{orcidlogocol}{HTML}{A6CE39}
\tikzset{
	orcidlogo/.pic={
		\fill[orcidlogocol] svg{M256,128c0,70.7-57.3,128-128,128C57.3,256,0,198.7,0,128C0,57.3,57.3,0,128,0C198.7,0,256,57.3,256,128z};
		\fill[white] svg{M86.3,186.2H70.9V79.1h15.4v48.4V186.2z}
		svg{M108.9,79.1h41.6c39.6,0,57,28.3,57,53.6c0,27.5-21.5,53.6-56.8,53.6h-41.8V79.1z M124.3,172.4h24.5c34.9,0,42.9-26.5,42.9-39.7c0-21.5-13.7-39.7-43.7-39.7h-23.7V172.4z}
		svg{M88.7,56.8c0,5.5-4.5,10.1-10.1,10.1c-5.6,0-10.1-4.6-10.1-10.1c0-5.6,4.5-10.1,10.1-10.1C84.2,46.7,88.7,51.3,88.7,56.8z};
	}
}
 \allowdisplaybreaks[4]
\newcommand\orcidicon[1]{\href{https://orcid.org/#1}{\mbox{\scalerel*{
				\begin{tikzpicture}[yscale=-1,transform shape]
					\pic{orcidlogo};
				\end{tikzpicture}
			}{|}}}}
\tikzset{bold/.style={color=blue, line width=2pt}}
\tikzset{redop/.style={circle,fill=red}}
\tikzset{blueop/.style={circle,fill=blue}}

\usepackage{datetime}
\usepackage{comment}
\usepackage{lineno}
\usepackage{xcolor}

\newcommand{\mT}{\mathsf{T}}
\newcommand{\mt}{\mathsf{T}}

\newcommand{\divides}{\mathrel{|}}

\newcommand{\ZZ}{\mathbb{Z}}

\newcommand{\rad}{\mathrm{rad}}

\renewcommand{\i}{\mathrm{i}}

\allowdisplaybreaks[4]

\graphicspath{{figure1/}}

\begin{document}

 \title{Infinite-component $BF$ topological field theory: connection of fracton order, Toeplitz braiding, and  non-Hermitian Amplification}

\author{Bo-Xi Li}
\affiliation{School of Physics, State Key Laboratory of Optoelectronic Materials and Technologies, and Guangdong Provincial Key Laboratory of Magnetoelectric Physics and Devices, Sun Yat-sen University, Guangzhou, 510275, China}

\author{Peng Ye\orcidicon{0000-0002-6251-677X}}
\email{yepeng5@mail.sysu.edu.cn}
\affiliation{School of Physics, State Key Laboratory of Optoelectronic Materials and Technologies, and Guangdong Provincial Key Laboratory of Magnetoelectric Physics and Devices, Sun Yat-sen University, Guangzhou, 510275, China}

\date{\today}
 \begin{abstract}
Building on the infinite-component Chern--Simons theory of three-dimensional fracton phases by Ma \textit{et al.}~[\href{https://doi.org/10.1103/PhysRevB.105.195124}{Phys.\ Rev.\ B~\textbf{105}, 195124 (2022)}] and the Toeplitz braiding of anyons by Li \textit{et al.}~[\href{https://doi.org/10.1103/PhysRevB.110.205108}{Phys.\ Rev.\ B~\textbf{110}, 205108 (2024)}], we show that stacking $(3+1)$D $BF$ topological field theories, which serve as low-energy effective descriptions of a class of three-dimensional topological orders, along a fourth spatial direction gives rise to an exotic class of four-dimensional fracton phases. Their low-energy physics is governed by a new field-theoretic framework, namely \textit{infinite-component $BF$} (i$BF$) \textit{theories}, characterized by asymmetric integer Toeplitz $K$ matrices. Under open boundary conditions along the stacking direction, i$BF$ theories with properly chosen $K$ matrices exhibit a striking phenomenon termed \textit{Toeplitz particle--loop braiding}, where a particle and a loop placed on opposite three-dimensional boundaries acquire a strongly oscillating yet robustly nonvanishing braiding phase even at infinite separation. This nonlocal braiding admits a geometric interpretation: adiabatically transporting the particle induces a winding boundary trajectory on the opposite boundary that encircles the loop. We show that this robustness originates from boundary zero singular modes (ZSMs) of Toeplitz $K$ matrices revealed by singular value decomposition, rather than from boundary zero eigenmodes responsible for previously known Toeplitz braiding of anyons, and that the same ZSM mechanism also underlies directional amplification in the rapidly developing field of non-Hermitian physics. We analytically and numerically study representative i$BF$ theories with Hatano--Nelson--type and non-Hermitian Su--Schrieffer--Heeger--type $K$ matrices, establishing a universal correspondence between ZSMs and Toeplitz particle--loop braiding. Our results identify boundary zero singular modes as the operative mechanism behind Toeplitz particle--loop braiding and establish infinite-component $BF$ theory as a predictive framework for higher-dimensional fracton topological orders.
\end{abstract}

\maketitle

\tableofcontents

\section{Introduction\label{intro}}
\subsection{Infinite-component Chern-Simons theory, anyonic Toeplitz braiding, and topological edge states}

Recently, \textit{fracton topological order} (FTO) has emerged as a distinctive
frontier in quantum many-body systems~\cite{fracton,fractonorder1,fractonorder2,fractonorder3,fracton16},
driving advances in condensed matter and quantum physics~\cite{PhysRevLett.129.230502,fracton3,fracton2,fracton1,fracton19,
fracton4,fracton17,hu2025,fracton18,fracton51,fracton52,fracton27,fracton53,
fracton54,fracton15}.
Solvable realizations of FTO, known as fracton codes, include the 3D\footnote{In this paper, ``$(n+1)$D'' refers to $(n+1)$-dimensional spacetime with $n$-dimensional real space. When referring specifically to spatial dimensions, we use ``$n$D''.} X-cube model and Haah’s
code~\cite{fractonorder2,fractonorder3,fracton16}.
In arbitrary dimensions, generalizations called ``tetradigit stabilizer codes''
have been proposed~\cite{hu2025,fracton18,fracton51,fracton52}.
Their excitations—collectively referred to as \textit{fractonic excitations}—exhibit
restricted mobility: point-like fractons are immobile, lineons move only along
lines, and planeons are confined to planes.
This sharply contrasts with liquid-like topological orders such as the toric code,
where anyons propagate freely via string operators.
Studies of FTO have further inspired subsystem symmetry-protected topological
phases~\cite{SSPT9,SSPT3,SSPT4,SSPT5,SSPT6,SSPT7,SSPT10,SSPT11,SSPT12,SSPT8,
zhang2025programmable,SSPT1,SSPT2} and many-body systems with conserved multipole
moments~\cite{fracton58,fracton59,Fractongauge}.
The former generalize global symmetries to lower-dimensional subsystems; the latter
impose strong kinematic constraints, yielding unconventional quantum
phenomena~\cite{fracton55,fracton56,fracton57,fracton60,fracton20,fracton26,
fracton5,diffusionofhigher-moment,Fractonicsuperfluids1,Fractonicsuperfluids2,
wanghanxie,NSofFractonicsuperfluids,fracton23,Fractonicsuperfluidsdefect,
reviewofFractonicsuperfluids,fracton28,fracton7,fracton21,fracton14,fracton22,
fracton30,Lifshitzduality,fracton11,HMWT,DBHM,DBHM2,fracton31,fracton35,
fracton44,fracton45,fracton6,fracton29,fracton24,fracton34,fracton36,
fracton37,fracton33,fracton38,fracton39,fracton40,fracton9,fracton41,fracton32,
fracton43,fracton48,fracton8,fracton42,fracton46,fracton25,fracton10,fracton12,
fracton49,fracton50,fracton47,rank_two_TC,rank_two_TC_Z_N,effective_2d_fracton,
2024PhRvB.109t5146D,2024arXiv240612962P}.

A large subclass of FTOs exhibits a foliation structure, allowing them to be
viewed as stacks of coupled 2D topological orders.
This perspective provides a natural route to field-theoretical descriptions of
fracton phases. Recent studies have developed stacking-based continuum or effective
field theories~\cite{shimamura2024anomaly,godbillon2025,Ebisu2024multipole,
spieler2023exotic,Ohmori2023,hsin2023gapped,fracton27,ma2022fractonic,
chen2022gapless,chen2023ground,wu2023transition,sullivan2021weak,
sullivan2021planar,levin2009gapless,li2024}.
Among these, the infinite-component Chern--Simons (iCS) theory—constructed by
stacking multicomponent Chern--Simons layers—has proven particularly promising~\cite{ma2022fractonic,chen2022gapless,chen2023ground,wu2023transition,
sullivan2021weak,sullivan2021planar,levin2009gapless,li2024}.
Before proceeding, we briefly recall the basic structure of the $(2+1)$D
multicomponent Chern--Simons theory, which serves as a prototypical topological
field theory capturing the low-energy physics of Abelian topological orders.

In $(2+1)$D multicomponent Chern--Simons theory, topological quantities such as
braiding statistics, ground-state degeneracy on a torus, and chiral central charge
are encoded in the symmetric integer $K$ matrix~\cite{wen1992classification,wen2004quantum}.
This framework covers many topologically ordered phases, including multilayer
fractional quantum Hall states and certain spin liquids~\cite{wen2004quantum}.
It also provides a systematic method for classifying symmetry-enriched (SET) and
symmetry-protected (SPT) phases in 2D~\cite{lu2012theory,lu2016classification,YW12,Ye14b,hung2013kmatrix,
PhysRevB.93.115136,meng2014topological}.
The Abelian Chern--Simons theory considered here follows a hydrodynamical approach,
distinct from the flux-attachment formulation~\cite{PhysRevLett.62.82,doi:10.1142/S0217979292000037,
composite_Fradkin,PhysRevB.94.115104}.

To construct 3D FTOs, we treat each 2D ($xy$ plane) Abelian topological order
described by $(2+1)$D multicomponent Chern--Simons theory as a 2D ``layer'' and
stack them along the $z$ direction, coupling adjacent layers via additional
Chern--Simons terms that preserve translational invariance.
This yields a multicomponent Chern--Simons theory whose topological data are
encoded in a symmetric integer block-Toeplitz ($K$) matrix~\cite{CIT-006,li2024}.
The action of such a theory is
\begin{equation}
   \!\!\! S=\int \frac{K_{I,J}}{4\pi} a^I \wedge da^J,\,
    \scalebox{0.9}{$
    K=\left(
    \begin{matrix}
        A & B^\mathsf{T} & & & & \\
        B & A & B^\mathsf{T} & & & \\
          & B & A & B^\mathsf{T} & & \\
          &   & \ddots & \ddots & \ddots & \\
          &   &        & B & A & B^\mathsf{T} \\
          &   &        &   & B & A \\
    \end{matrix}
    \right)
    $}.
\end{equation}
Here $a^{I,J}$ ($I,J=1,\ldots,N$) are $U(1)$ compact gauge fields, and implicit
summation is assumed over these indices, which also label positions along $z$.
The absence of lower-left and upper-right blocks in $K$ indicates open boundary
conditions (OBC) along the stacking direction.
In the limit $N\to\infty$, this construction yields the iCS theory.
The system boundaries fall into two classes: the $xy$ surfaces of each layer and
the $z$ boundaries along the stacking direction.
While the $xy$ boundaries are conventional and can be described by stacked conformal
field theories, the novel $z$ boundaries were only understood after the discovery
of Toeplitz braiding in Ref.~\cite{li2024}.

Specifically, under OBC in the $z$ direction, boundary zero modes of the $K$ matrix,
together with noninteger elements in the upper-right and lower-left corners of
$K^{-1}$, generate nonlocal braiding statistics between particles on opposite
$z$ boundaries. Such exotic braiding statistics is demonstrated in Figs.~5 and 6
of Ref.~\cite{li2024}.
This phenomenon, termed ``Toeplitz braiding'', reveals a mathematical structure
closely resembling the boundary theory of topological insulators and superconductors
in free-fermion systems~\cite{qi2011topological}, establishing a link between
strongly and weakly correlated systems.

\subsection{Infinite-component $BF$ field theory, particle-loop Toeplitz braiding and boundary zero singular modes\label{sectionIB}}

\begin{figure}
	\centering
    \includegraphics[width=\columnwidth]{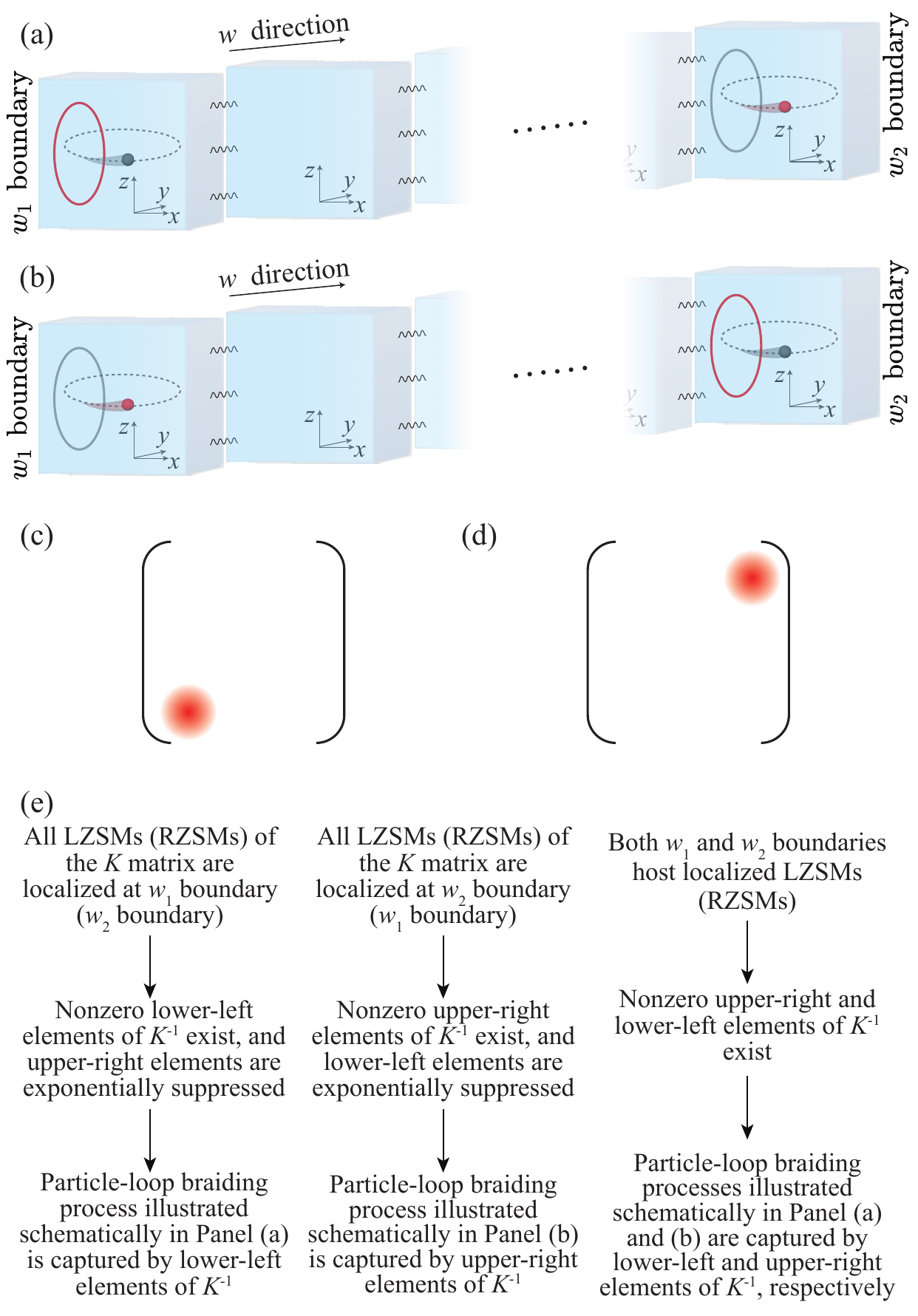}
    \caption{
Illustration of stacking and Toeplitz particle--loop braiding in i$BF$ theories. Each so-called ``layer'' is a genuine three-dimensional $BF$ theory defined on a three-torus and is schematically represented by a light-blue cube. These 3D layers are stacked along the $w$ direction (i.e., the extra spatial dimension) and coupled by interlayer $BF$ terms (wavy lines), giving rise to two $(3+1)$D boundaries labeled as $w_1$ and $w_2$. Panel (a) shows Toeplitz particle--loop braiding between a loop excitation (red) on the $w_1$ boundary and a particle excitation (red) on the $w_2$ boundary. A nontrivial braiding phase arises when the particle (red) is adiabatically transported such that its \emph{induced boundary trajectory} (blue) on the \emph{opposite} boundary---namely, the projected ``shadow'' of the particle (red) onto the $w_1$ boundary---winds around the loop (red). This process is encoded in the lower-left elements of $K^{-1}$, highlighted in panel (c). Panel (b) depicts the opposite configuration, encoded in the upper-right elements of $K^{-1}$ highlighted in panel (d). Panel (e) summarizes the correspondence between the boundary localization of left and right zero singular modes (LZSMs and RZSMs), the structure of $K^{-1}$, and the resulting Toeplitz particle--loop braiding. These summarized results are higher-dimensional generalization of Toeplitz braiding introduced in Ref.~\cite{li2024}.}
    \label{particleloopbraiding}
\end{figure}

Building on the above iCS framework and the concept of Toeplitz braiding between anyons,
this work takes a further step by extending the stacking construction from 2D to 3D
topological orders, thereby realizing 4D FTOs. Our primary goal is to explore the
resulting Toeplitz braiding between particles and loops within this higher-dimensional
framework, as shown in Fig.~\ref{particleloopbraiding}. In contrast to conventional particle--loop linking processes in three
dimensions, the braiding phases uncovered here arise from boundary-induced geometric
winding and are intrinsically nonlocal along the stacking direction. As shown in Fig.~\ref{particleloopbraiding}, although the particle and the loop reside
on opposite $(3+1)$D boundaries and may be separated by an arbitrarily large distance
along the stacking ($w$) direction, adiabatically transporting the particle does not remain
invisible to the loop. Instead, due to the interlayer $BF$ couplings (to be introduced shortly), the motion
of the particle induces an effective boundary trajectory on the opposite boundary,
depicted by the blue path in Fig.~\ref{particleloopbraiding}(a). When this induced
boundary trajectory adiabatically winds around the loop excitation, a nontrivial braiding phase
is accumulated. Crucially, the robustness of this phase depends only on the winding topology of the
induced boundary trajectory and is independent of the separation between the particle
and the loop along the $w$ direction. From a field-theoretic perspective, in this work, we introduce the ``infinite-component BF field theory'' (i$BF$) to describe the above Toeplitz braiding. 

The usual $(3+1)$D multicomponent $BF$ theories~\cite{horowitz_quantum_1990,hansson_superconductors_2004}
provide a versatile framework for describing 3D topological orders.
Analogous to the $(2+1)$D Chern--Simons description of 2D systems, when pointlike and
loop excitations are represented as charges and fluxes of an Abelian finite gauge group,
the $BF$ theory and its twisted variants~\cite{PhysRevB.94.115104,Moy_Fradkin2023,YW13a,
2016arXiv161209298P,PhysRevB.99.235137,YeGu2015,ypdw,PhysRevLett.121.061601,
zhang_compatible_2021,bti2,bti6} form a unified field-theoretic scheme for capturing
3D topological orders and \textit{gauged} SPT phases.
A finite set of canonical $BF$ terms $BdA$ (with 2-form $B$ and 1-form $A$ fields), together with
twisted terms such as $AAdA$, $AAAA$, $AAB$, $BB$, and the $\theta$-term $dAdA$,
encodes rich topological data—including particle--loop, multiloop, and Borromean rings
braiding~\cite{hansson_superconductors_2004,PRESKILL199050,PhysRevLett.62.1071,
PhysRevLett.62.1221,ALFORD1992251,wang_levin1,PhysRevLett.114.031601,
2016arXiv161209298P,string4,jian_qi_14,string5,PhysRevX.6.021015,Tiwari:2016aa,
corbodism3,string6,3loop_ryu,PhysRevLett.121.061601}, emergent fermionic
statistics~\cite{Kapustin:2014gua,bti2,PhysRevB.99.235137,Zhang2023Continuum},
and topological responses~\cite{qi2011topological,lapa17,YW13a,PhysRevB.94.115104,Ye:2017aa,
bti6,RevModPhys.88.035001,PhysRevB.99.205120}.
Refs.~\cite{Ning2018prb,ye16_set,2016arXiv161008645Y} further showed how global
symmetries fractionalize on loop excitations, introducing ``mixed three-loop braiding'',
which underlies the classification of higher-dimensional SET phases.
These braiding processes obey stringent consistency conditions~\cite{zhang_compatible_2021}
ensuring gauge invariance and allowing for non-Abelian fusion and shrinking rules even
when loop charges remain Abelian~\cite{PhysRevLett.121.061601,Zhang2023fusion}.
Building on this foundation, a diagrammatic representation for higher-dimensional
topological orders—including 4d cases~\cite{Huang2023}—was recently developed in
Ref.~\cite{2024arXiv240519077H}.
Moreover, recent studies have explored novel self-statistics of fully immobile excitations in FTOs \cite{PhysRevLett.132.016604}, demonstrating that braiding statistics in FTOs can differ from those in liquid-like topological orders with fully mobile excitations. 

To this end, we generalize the above established multicomponent $BF$ framework to an  i$BF$ theory—to describe 4d FTOs and Toeplitz braiding phenomena.
In analogy to the iCS construction, we couple multicomponent $(3+1)$D $BF$ theories to
formulate an effective field theory for 4d FTOs:
$S=\int \frac{A_{I,J}}{2\pi}\, b^I \wedge d a^J$.
Stacking $N$ layers of such $(3+1)$D field theories and coupling neighboring 3D ``layers''  via
interlayer $BF$ terms while preserving translational symmetry yields a multicomponent
$BF$ theory characterized by a Toeplitz $K$ matrix with integer elements:
\begin{gather}
	S=\int \frac{K_{I,J}}{2\pi}b^I \wedge d a^J,\, 
	 \scalebox{0.9}{$ K = \left( \begin{matrix}
   A & C & & & & \\
   B & A & C & & & \\
     & B & A & C & & \\
     &   & \ddots& \ddots & \ddots & \\
     &   &       & B & A & C \\
     &   &       &   & B & A \\
    \end{matrix} \right)$}.\label{kmatrix}
\end{gather}
Here $N$ denotes the number of layers, also referred to as the ``system size''.
Taking $N\!\to\!\infty$ defines the \textit{infinite-component $BF$ theory}, i.e.,  i$BF$ theory.
The stacking process is illustrated in Figs.~\ref{particleloopbraiding}(a) and (b),
where OBC is imposed along the $w$ direction. Each blue cube represents a 3D ``layer''
of $BF$ theory on a three-torus where periodic boundary condition (PBC) is applied along
$x$, $y$, and $z$, and the wavy lines denote interlayer $BF$ couplings.
A detailed discussion of particle--loop braiding and the construction of i$BF$ theory is
given in Sec.~\ref{sectionmbf}.

Given an i$BF$ theory with nearest-neighbor coupling, a natural question arises:
\emph{under what conditions does nonlocal braiding along the stacking direction emerge?
Is there an analog of boundary-mode-induced nonlocal braiding—i.e., Toeplitz braiding—
along the stacking direction?} To address this, we classify i$BF$ theories into two types.
The first involves symmetric $K$ matrices, where $C=B^\mt$ in Eq.~(\ref{kmatrix}); in this
case, the Toeplitz braiding analysis parallels that of iCS theories~\cite{li2024}, except
that the particle--particle braiding of Ref.~\cite{li2024} now corresponds to particle--loop
braiding in the $BF$ context.

As shown in Sec.~\ref{sectionIII}, when $K$ becomes asymmetric—sharing the same mathematical
structure as the one-dimensional Hatano--Nelson (HN)~\cite{PhysRevLett.77.570} or non-Hermitian
Su--Schrieffer--Heeger (nSSH)\footnote{In this paper, SSH only stands for \textit{Hermitian} Su--Schrieffer--Heeger.}~\cite{PhysRevA.89.062102,PhysRevB.97.045106} Hamiltonians—the singular value
decomposition (SVD)~\cite{horn2012matrix} of $K$ reveals \emph{boundary zero singular modes}
(ZSMs)\footnote{By ZSMs, we always mean modes from boundaries to avoid lengthy abbreviation
such as BZSM.\label{footnote_zsm}} that give rise to nonlocal braiding along the stacking
($w$) direction. This phenomenon, also termed \textit{Toeplitz braiding}, cannot be captured
by boundary zero eigenmodes, emphasizing the necessity of the SVD perspective.

When all ``left'' ZSMs (LZSMs) localize on one $w$ boundary and ``right'' ZSMs (RZSMs) on the
other,\footnote{Here the ``left'' and the ``right'' label internal sectors of the SVD rather
than spatial sides.} either the lower-left or upper-right elements of $K^{-1}$ decay
exponentially with system size. Consequently, nontrivial braiding arises when particles and
loops occupy specific opposite $w$ boundaries, as illustrated in
Figs.~\ref{particleloopbraiding}(a) and \ref{particleloopbraiding}(b).
The corresponding braiding phases are encoded in the lower-left or upper-right elements of
$K^{-1}$ [Figs.~\ref{particleloopbraiding}(c) and \ref{particleloopbraiding}(d)].
In either case, exchanging the particle and loop positions rapidly suppresses the braiding
phase upon increasing the size of the $w$ direction.
In particular, i$BF$ theories with $K$ matrices resembling the nSSH Hamiltonians can exhibit
nontrivial Toeplitz braiding in both configurations shown schematically in
Figs.~\ref{particleloopbraiding}(a) and~\ref{particleloopbraiding}(b), as the $K$ matrices may
host two LZSMs (RZSMs) that localize at distinct $w$ boundaries. The relation between the
locations of the LZSMs and RZSMs, the patterns of $K^{-1}$, and the resulting braiding
statistics is summarized in Fig.~\ref{particleloopbraiding}(e).

Interestingly, the same mathematical structure also appears in non-Hermitian directional
amplification in driven–dissipative systems~\cite{porras2019topological,PhysRevA.103.033513,
brunelli2023restoration,PhysRevB.103.L241408}. When ZSMs arise in such non-Hermitian
Hamiltonians, driven–dissipative cavity arrays may exhibit directional amplification of
coherent inputs, with the gain growing exponentially with system size, whereas reversing the
input direction markedly suppresses the effect.

The paper is organized as follows.
Section~\ref{sectionmbf} establishes the braiding structure of multicomponent $BF$ theories and
introduces the stacking construction that gives rise to i$BF$ field theories, emphasizing why
non-diagonal Toeplitz $K$ matrices are essential. Sections~\ref{sectionIII} and
\ref{sectionIIIB} uncover Toeplitz braiding in i$BF$ theories whose $K$ matrices inherit the
mathematical structures of the HN and nSSH Hamiltonians, respectively. These sections show how
SVD sharply exposes boundary zero singular modes and reveal a direct analogy to non-Hermitian
directional amplification. Section~\ref{conclusion} closes with a brief summary and outlook.

\section{Braiding  in i$BF$ theories\label{sectionmbf}}
 \subsection{Multicomponent $BF$ theories}
 
In 3D Abelian topological orders, the topological excitations include both particles and loops; the $BF$ theory naturally captures the associated Aharonov--Bohm particle--loop braiding phase. 
The partition function of a multicomponent $BF$ theory\footnote{While a single-component 
$BF$ theory usually suffices for a given $3{+}1$D spacetime, the multicomponent 
generalization is essential for our construction of Toeplitz braiding. A related example 
is the charge--loop excitation symmetry in Ref.~\cite{PhysRevB.94.115104}.} is
\begin{equation}
    Z \;=\; \int \mathcal{D}b^I\,\mathcal{D}a^J\, e^{\,\mathrm{i} S_{BF}},\,\, 
    S_{BF} \;=\; \int \frac{K_{I,J}}{2\pi}\, b^I \wedge d a^J .
\end{equation}
Here $K$ is a general $N\times N$ matrix (not necessarily Toeplitz), whose constraints will be 
determined below. Repeated indices $I,J$ are implicitly summed over. The fields $b^I$ and $a^J$ 
($I,J = 1,\ldots,N$) are compact $U(1)$ 2-form and 1-form gauge fields obeying the Dirac 
quantization conditions
\begin{equation}
    \oint db^I \in 2\pi\mathbb{Z},
    \qquad 
    \oint da^I \in 2\pi\mathbb{Z}.
\end{equation}

Their gauge transformations are $b^I\!\to\! b^I+d\beta^I$ and $a^I\!\to\! a^I+d\alpha^I$, where $\beta^I$ (1-form) and $\alpha^I$ (0-form) are compact gauge parameters obeying $\oint d\beta^I\!\in\!2\pi\mathbb{Z}$ and $\oint d\alpha^I\!\in\!2\pi\mathbb{Z}$. Under these transformations, the action changes only by a boundary term: $S_{BF}\!\to\!S_{BF}+\frac{K_{I,J}}{2\pi}\!\int d\beta^I\!\wedge\! da^J$. Placing the theory on $S^2\!\times\!S^2$ and decomposing the integral along the two spheres immediately requires $K_{I,J}\!\in\!\mathbb{Z}$, ensuring gauge invariance of the partition function.

The observables are Wilson operators
\begin{align}
	W(\gamma,\omega)
	&=\exp\!\left[\mathrm{i}\!\int_{\omega} L_I b^I+\mathrm{i}\!\int_\gamma N_I a^I \right]\nonumber\\
	&=\exp\!\left[\mathrm{i}\!\int L_I b^I\!\wedge\!\delta\omega
	+\mathrm{i}\!\int N_I a^I\!\wedge\!\delta\gamma \right],
\end{align}
where $\gamma$ is a closed 1D loop and $\omega$ a closed 2D surface. The delta forms $\delta\omega$ and $\delta\gamma$ are supported only on these manifolds. Gauge invariance of $W(\gamma,\omega)$ requires $L_I,N_I\in\mathbb{Z}$. A topological excitation is thus either a particle labeled by $\mathbf{N}=(N_1,\ldots,N_N)^{\mT}$ or a closed loop labeled by $\mathbf{L}=(L_1,\ldots,L_N)^{\mT}$, whose worldline and worldsheet are $\gamma$ and $\omega$, respectively. The expectation value reads
\begin{equation}
	\langle W(\gamma,\omega)\rangle
	= \frac{1}{Z}\!\int\! \mathcal D b^I \mathcal D a^J\, W(\gamma,\omega)\, e^{\,\mathrm{i} S_{BF}}.
\end{equation}
where $Z$ is the partition function for the purpose of normalization.   Integrating out $b^I$ gives $L_I\delta\omega+\frac{K_{I,J}}{2\pi}d a^J=0$, so that $da^I=-2\pi (K^{-1})_{I,J} L_J \delta\omega$. Under a proper gauge choice, the solution is $a^I=-2\pi (K^{-1})_{I,J}L_J \delta\Omega$, where $\Omega$ is a Seifert hypersurface bounded by $\omega$ ($\partial\Omega=\omega$). Hence,
\begin{equation}
	\langle W(\gamma,\omega)\rangle
	=\exp\!\left[-2\pi \mathrm{i}\, N_I (K^{-1})_{I,J} L_J
	\!\int \delta\Omega\!\wedge\!\delta\gamma \right].
	\label{wilsonbf1}
\end{equation}
The integral $\int \delta\Omega\!\wedge\!\delta\gamma$ represents the intersection number between  $\Omega$ and $\gamma$, which is the linking number between $\omega$ and $\gamma$ by definition. Thus, the Aharonov–Bohm phase between a particle and a loop is~\cite{PhysRevB.94.115104}
\begin{equation}
	2\pi N_I (K^{-1})_{I,J} L_J
	=2\pi\, \mathbf{N}^\mT K^{-1} \mathbf{L}.
	\label{braidingphaseexp}
\end{equation}
This shows that particle–loop braiding phases are encoded in $K^{-1}$, directly analogous to particle–particle braiding in (2+1)D multicomponent Chern–Simons theories.

\subsection{Construction of infinite-component $BF$ theories}

As outlined in Sec.~\ref{intro}, the i$BF$ theory arises from coupling multicomponent (3+1)D $BF$ theories of the form
$S=\int \frac{A_{I,J}}{2\pi}\, b^I\wedge d a^J$, where $b^I$ and $a^J$ denote compact $U(1)$ 2-form and 1-form gauge fields, respectively.
Here $A$ is a square integer matrix, and repeated indices $I,J$ are summed over.
By stacking $N$ layers of 3D topological orders described by these $BF$ theories and introducing interlayer $BF$ couplings that preserve translational symmetry along the stacking direction, we obtain a multicomponent $BF$ theory characterized by an integer Toeplitz $K$ matrix that encodes the braiding statistics of topological excitations, as given in Eq.~(\ref{kmatrix}).

Taking the number of layers $N\!\to\!\infty$ defines the \emph{infinite-component $BF$ theory} (i$BF$ theory).  In the upcoming discussions, taking the thermodynamic limit refers specifically to sending $N$ to infinity, as each individual layer is already in its own thermodynamic limit.
The analysis in Sec.~\ref{sectionIII} follows this framework: first compute $K^{-1}$ for finite $N$, then study the $N\!\to\!\infty$ limit to extract asymptotic braiding behavior. 
In what follows, we denote the stacking direction as the $w$ axis and label the two boundaries as the $w_1$ and $w_2$ boundaries.  
A loop or particle excitation $(1,0,\ldots,0)^{\mT}$ is assigned to the $w_1$ boundary, while $(0,0,\ldots,1)^{\mT}$ lies on the $w_2$ boundary.  
\textit{As a result, the indices of the components of the charge vectors $\mathbf{N}$ and $\mathbf{L}$ carry the meaning of $w$ coordinates.}

Fig.~\ref{particleloopbraiding} schematically illustrates this stacking and particle–loop braiding process: each layer is a 3-torus [blue cubes in Figs.~\ref{particleloopbraiding}(a) and \ref{particleloopbraiding}(b)], the wavy lines represent interlayer $BF$ terms, and the stacking direction is the $w$ axis.  
Here, we introduce a useful quantity, $\Theta_{I,J}$, which encodes the spatial distribution of particle--loop braiding phases along the $w$ direction:
\begin{gather}
    \Theta_{I,J} = 2\pi (K^{-1})_{I,J}.\label{thetaij}
\end{gather}
With proper $2\pi$ shift, $\Theta_{I,J}$ is defined in $(-\pi,\pi]$ unless otherwise specified. More precisely, \textit{$\Theta_{I,J}$ is the braiding phase between a particle carrying unit charge located at $w$ coordinate $I$ and a loop carrying unit charge located at $w$ coordinate $J$.} The particle and loop vectors are given respectively:
\begin{subequations}
    \begin{gather}
\mathbf{N}=(0,0,\cdots,1,\cdots,0,0)^{\mT}\, ,\\
\mathbf{L}=(0,0,\cdots,1,\cdots,0,0)^{\mT}\, .
\end{gather}
\end{subequations}
where the two entries $1$'s are respectively located at the $I$th and $J$th components of $\mathbf{N}$ and $\mathbf{L}$ i.e., $N_{I'}=\delta_{I,I'},L_{J'}=\delta_{J,J'}$. 
Accordingly, the braiding between a loop at the $w_1$ boundary and a particle at the $w_2$ boundary [Fig.~\ref{particleloopbraiding}(a)] is encoded in the lower-left elements of $K^{-1}$ as well as the matrix $\Theta$ [Fig.~\ref{particleloopbraiding}(c)], while the reverse configuration [Fig.~\ref{particleloopbraiding}(b)] corresponds to the upper-right elements [Fig.~\ref{particleloopbraiding}(d)].

Before analyzing specific i$BF$ theories, we emphasize the necessity of considering non-diagonal Toeplitz $K$ matrices in Eq.~(\ref{kmatrix}).
Conventional multicomponent $BF$ theories possess a ``relabeling redundancy''~\cite{PhysRevB.94.115104}: if two different $K$ matrices denoted as $K$ and $K'$ of the same size $N\times N$, are related by two independent $GL(N,\mathbb{Z})$ transformations $\Omega,W$ ($|\det \Omega|=|\det W|=1$, $W_{I,J},\Omega_{I,J}\in\ZZ$) as $K'=\Omega K W^\mt$, and the particle and loop charge vectors transform as $\mathbf{N}'=W \mathbf{N}$ and $\mathbf{L}'=\Omega\mathbf{L}$, then we obtain identical braiding phases, $2\pi {\mathbf{N}'}^{\mT} {{K}'}^{-1} \mathbf{L}'=2\pi \mathbf{N}^{\mT} {K}^{-1}\mathbf{L}$.   
Since any integer matrix can be brought to its Smith normal form via $GL(N,\mathbb{Z})$ transformations, one can normally decouple mutual $BF$ terms by such diagonalization and study the diagonalized $K$ matrix.
However,  by noting that $I, J$ indices now  label the $w$ coordinates for i$BF$ theories, preserving locality along the stacking direction restricts the allowed $GL(N,\mathbb{Z})$ transformations for Toeplitz $K$ matrices [Eq.~(\ref{kmatrix})]: $\Omega$ and $W$ must be quasi-diagonal, with nonzero elements only clustered near the main diagonal.  

To conclude this section, we point out   the above diagonalization transformations on the Toeplitz matrix $K$ via $GL(N,\mathbb{Z})$  cannot be found for the nontrivial Toeplitz braiding situation where $K^{-1}$ possesses nonzero elements in either the upper-right or lower-left corners (see Fig.~\ref{particleloopbraiding}). 
This conclusion can be straightforwardly proved via \textit{reductio ad absurdum}:
if a Toeplitz $K$ is brought to its Smith normal form $\Lambda$ (with nonzero elements only on the main diagonal) by quasi-diagonal $\Omega$ and $W$, i.e.,  $\Lambda=\Omega K W^\mt$ or equivalently $K^{-1}=W^\mt \Lambda^{-1}\Omega$. Obviously, due to the above-mentioned properties of $W$ and $\Omega$, $K^{-1}$ has nonzero elements only clustered near the main diagonal. This, in turn, forbids any possible nonzero elements in the upper-right or lower-left corners, contradicting the existence of the nonzero upper-right and lower-left elements responsible for nontrivial Toeplitz braiding as shown in Fig.~\ref{particleloopbraiding}.
Hence, non-diagonal Toeplitz $K$ matrices are essential for the existence of the nontrivial Toeplitz braiding of i$BF$ theories.

 \section{i$BF$ theory with Hatano-Nelson-type $K$ matrix\label{sectionIII}}

As described in Sec.~\ref{intro}, we classify the $K$ matrices of i$BF$ theories into two types.
The first type consists of symmetric $K$ matrices, where $C = B^\mt$ in Eq.~(\ref{kmatrix}).
In this case, the analysis of Toeplitz braiding parallels that in iCS theories~\cite{li2024}, 
except that the particle–particle braiding discussed in Ref.~\cite{li2024} is now reinterpreted as particle–loop braiding in the $BF$ framework.
When $K$ is asymmetric, however, the particle–loop braiding phase is no longer invariant under exchanging the $w$ coordinates of the particle and the loop, because $K^{-1}$ is also asymmetric.
As we will show, the presence of ZSMs leads to nonlocal braiding along the $w$ direction. All detailed relations are summarized in Fig.~\ref{particleloopbraiding}.

In this section and following section, we demonstrate these features using two representative examples:
the i$BF$ theory with a HN–type $K$ matrix and that with a nSSH–type $K$ matrix.  
Both of them arise from coupling multicomponent $BF$ theories [Eq.~(\ref{kmatrix})] with either $A=n$ or $A=\begin{pmatrix}
	A_{11} & A_{12} \\[2pt]
	A_{21} & A_{22}
\end{pmatrix},\qquad \det A\neq 0.$
For a $2\times2$ integer matrix $A$ of this form, its Smith normal form is $
\begin{pmatrix}
	d_1 & 0 \\[2pt] 0 & d_2
\end{pmatrix}$, 
where $d_1=\gcd(A_{11},A_{12},A_{21},A_{22})$ and 
$d_2=|A_{11}A_{22}-A_{12}A_{21}|/\gcd(A_{11},A_{12},A_{21},A_{22})$~\cite{dummit2004abstract}.  
Hence, the corresponding $BF$ theory effectively describes a $\ZZ_{d_1}\times \ZZ_{d_2}$ gauge theory.

\subsection{Toeplitz braiding and ZSMs--A concrete example}

\begin{figure}[htbp]
    \centering
    \includegraphics[width=\linewidth]{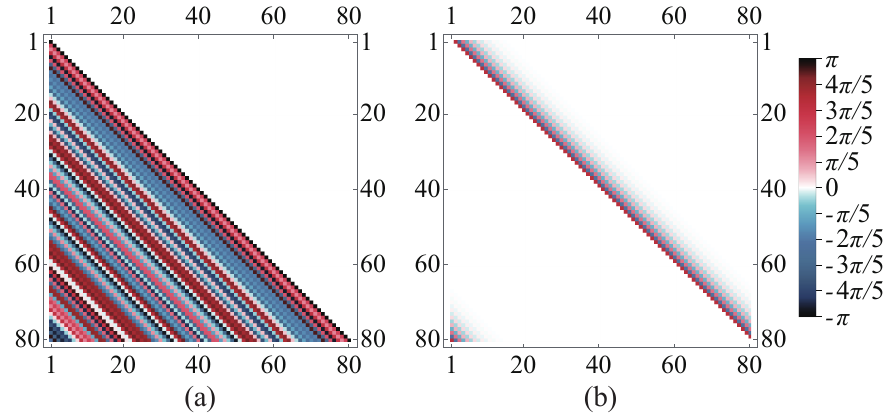}
    \caption{Braiding phase $\Theta_{I,J}=2\pi (K_{\mathrm{HN}}^{-1})_{I,J}$ (a) and $\Theta_{I,J}=2\pi (K_{\mathrm{HN,PBC}}^{-1})_{I,J}$ (b) encoded in $K_{\mathrm{HN}}$ and $K_{\mathrm{HN,PBC}}$ specified by $n=2,b=3$, where we select the system size $N=80$ for illustrative purpose. The lower-left nonzero elements in Panel (b) corresponds to the braiding phases $\Theta_{I,J}$ between test particles and loops with $w$ coordinates $I\sim N$, $J\sim 1$, which are close in the $w$ direction due to PBC.}
    \label{pbcics}
\end{figure}

We use a concrete example—i$BF$ constructed by coupling $\ZZ_{n}$ topological orders—to demonstrate the correspondence between ZSMs and nonlocal braiding statistics in stacking direction. 
  Before taking the $N\rightarrow \infty$ limit, the $K$ matrix of the i$BF$ theory writes
 \begin{gather}
	K_{\mathrm{HN}} = \left( \begin{matrix}
    n &  & & & & \\
    b & n &  & & & \\
    & b & n &  & & \\
    & & \ddots & \ddots &  & \\
    & & & b& n &  \\
    & & & & b & n \\
    \end{matrix} \right).\label{kmatrixhatanonb}
\end{gather}
It is worth noting that the $K$ matrix in Eq.~(\ref{kmatrixhatanonb}) shares the same mathematical form with the Hamiltonian of Hatano-Nelson model \cite{PhysRevLett.77.570} 
\begin{gather}
	K_{\mathrm{HN}}' = \left( \begin{matrix}
    n & c & & & & \\
    b & n & c & & & \\
    & b & n & c & & \\
    & & \ddots & \ddots & \ddots & \\
    & & & b& n & c \\
    & & & & b & n \\
    \end{matrix} \right)\label{kmatrixhatano}
\end{gather}
when $c=0$.
The inverse of $K_{\mathrm{HN}}$ encodes the particle-loop braiding statistics, which reads
\begin{gather}
	(K_{\mathrm{HN}}^{-1})_{I,J}= \left\{
    \begin{array}{l}
         \frac{1}{n}\left(-\frac{b}{n}\right)^{I-J},\quad I\geqslant J; \\
         0,\quad I < J  .
    \end{array}\right.\label{hninverseele}
\end{gather}
If $|b|>|n|$ together with $n\, \mathrm{rad} (n) \nmid b$,\footnote{Here $\mathrm{rad}(n)$ denotes the product of the distinct prime factors of $n$. The condition $n\,\mathrm{rad}(n)\nmid b$ simply means that $b$ is not divisible by $n\,\mathrm{rad}(n)$, such that the braiding phase angle $\Theta_{I,J}$ defined in Eq.~(\ref{thetaij}) is not trivially equal to $0 \text{ mod } 2\pi$.} then a nontrivial braiding statistical phase between the particle and the loop arises. From the equation for braiding phase [Eq.~(\ref{braidingphaseexp})] together with the expression for $K_{\mathrm{HN}}^{-1}$, only when a particle is placed on $w_2$ boundary and a loop is placed on $w_1$ boundary, the nontrivial braiding phase manifests. For example, the mutual braiding phase between a particle $\mathbf N=(0,\ldots,0,1)^\mt$ and a loop $\mathbf L=(1,0,\ldots,0)^\mt$ is $2\pi\frac{1}{n}\left(-\frac{b}{n}\right)^{N-1}$. 
In contrast, if the positions of the particle and the loop are exchanged, then the nontrivial braiding statistical phase between them vanishes. For illustrative purpose, we depict the $\Theta_{I,J}$ matrix [defined in Eq.~(\ref{thetaij})]  for $K_{\mathrm{HN}}$ specified by $n=2,b=3$ with system size $N=80$ in Fig.~\ref{pbcics}(a), clearly demonstrating this nonlocal braiding feature in the $w$ direction.

To deepen our understanding of the origin of this nonlocal braiding statistics, we dig into the SVD of $K_{\mathrm{HN}}$ matrix.
The SVD of $K_{\mathrm{HN}}$ is written as
 $  K_{\mathrm{HN}} =\sum_{i=1}^N\sigma_i\mathbf u_i\mathbf v_i^\mt=\sigma_1\mathbf u_1 \mathbf v_1^\mt+\sigma_2\mathbf u_2 \mathbf v_2^\mt+\cdots$, where the left singular modes $\{\mathbf u_i\}$ and the right singular modes $\{\mathbf v_i\}$ are mutually orthonormal, i.e.,
$\mathbf u_i^\mt \mathbf u_j = \delta_{i,j}, ~ \mathbf v_i^\mt \mathbf v_j = \delta_{i,j}$, and $\{\sigma_i\}$ are the corresponding singular values.

If $|b|>|n|$, then $K_{\mathrm{HN}}$ possesses the following left singular mode $\mathbf u_1$ located at the $w_1$ boundary and right singular mode $\mathbf v_1$ located at the $w_2$ boundary when $N$ is sufficiently large:
  \begin{gather}
	\mathbf u_1= \sqrt{\frac{1-(\frac{n}{b})^2}{1-(\frac{n}{b})^{2N}}}\begin{pmatrix}
		1 & -\frac{n}{b} & \cdots & \left(-\frac{n}{b}\right)^{N-1}
	\end{pmatrix}^\mT,\\
	\mathbf v_1=  \sqrt{\frac{1-(\frac{n}{b})^2}{1-(\frac{n}{b})^{2N}}}\begin{pmatrix}
		\left(-\frac{n}{b}\right)^{N-1} & \left(-\frac{n}{b}\right)^{N-2} & \cdots & 1
	\end{pmatrix}^\mT.
\end{gather}
Detailed calculation is presented in Appendix \ref{detailedhn}.
  Moreover, the corresponding singular value $\sigma_1$  is obtained from
\begin{gather}
	\sigma_1=\mathbf u_1^\mT K_{\mathrm{HN}} \mathbf v_1 = \frac{1-(\frac{n}{b})^2}{1-(\frac{n}{b})^{2N}}n\left(\frac{-n}{b}\right)^{N-1}.
\end{gather}
The singular modes such as $\mathbf u_1$ and $\mathbf v_1$ that reside at the boundaries, decay exponentially into the bulk, and have singular values approaching zero in the thermodynamic limit $N\!\to\!\infty$ are referred to as ZSMs, as introduced in Sec.~\ref{sectionIB}. Left and right ZSMs are abbreviated LZSM and RZSM, respectively.\footnote{The ``left'' and ``right'' here refer to internal SVD sectors as is introduced in Sec.~\ref{sectionIB}, and are not related to the spatial distribution of the ZSMs (see Footnote \ref{footnote_zsm}). The LZSMs, RZSMs and the singular values discussed in this work may differ from the conventional definitions by a sign, which can be migrated among them without affecting the corresponding terms in SVD.}
 For large $N$, we can construct an approximate   matrix      $M_{\mathrm{HN}}$ from $\sigma_1$, $\mathbf u_1$ and $\mathbf v_1$ to model the large-size behavior of $K^{-1}$:
\begin{gather}
	M_{\mathrm{HN}}=\frac{1}{\sigma_1} \mathbf v_1 \mathbf u_1^\mT,\quad
	(M_{\mathrm{HN}})_{I,J}=\frac{1}{n}\left(-\frac{b}{n}\right)^{I-J}.
\end{gather}
$M_{\mathrm{HN}}$ represents the ZSM contribution to the inverse $K_{\mathrm{HN}}^{-1}=\sum_{i=1}^N \sigma_i^{-1} ~ \mathbf v_i\mathbf u_i^\mt=\sigma_1^{-1} ~ \mathbf v_1 \mathbf u_1^\mt+\sigma_2^{-1} ~ \mathbf v_2 \mathbf u_2^\mt+\cdots$.
We observe that $(M_{\mathrm{HN}})_{I,J}=(K_{\mathrm{HN}}^{-1})_{I,J}$ for $I\geqslant J$, which implies the braiding statistics between  particles placed on the $w_2$ boundary and loops placed on the $w_1$ boundary are indeed encoded in ZSMs and the corresponding exponentially small singular value.

Furthermore, to emphasize the crucial role of ZSMs in generating nonlocal braiding statistics, we impose PBC along the $w$ direction and isolate the bulk contribution to the inverse of $K_{\mathrm{HN}}$. Under PBC, the Hatano-Nelson-type $K$ matrix becomes 
\begin{gather}
	K_{\mathrm{HN,PBC}}=\left( \begin{matrix}
    n &  & & & & b \\
    b & n &  & & & \\
    & b & n &  & & \\
    & & \ddots & \ddots &  & \\
    & & & b& n &  \\
    & & & & b & n \\
    \end{matrix} \right),
\end{gather}
and the $w$ boundaries together with ZSMs are absent.
In the large $N$ limit, the inverse of $K_{\mathrm{HN,OBC}}$ is 
\begin{gather}
	(K_{\mathrm{HN,PBC}}^{-1})_{I,J}=\left\{\begin{array}{l}
		-\frac{1}{n}\left(-\frac{n}{b}\right)^{N-(I-J)},\quad I\geqslant J ;\\
		-\frac{1}{n}\left(-\frac{n}{b}\right)^{J-I},\quad I<J.
	\end{array}\right.
\end{gather}
The formula for $(K_{\mathrm{HN,PBC}}^{-1})_{I,J}$ indicates the mutual braiding phases $\Theta_{I,J}$ between test particles and loops which are well-separated along the $w$ direction become negligibly small. As an illustration, Fig.~\ref{pbcics}(b) displays $\Theta_{I,J}$ for $K_{\mathrm{HN,PBC}}$ with parameters $n=2$, $b=3$, and system size $N=80$. Thus, braiding statistics is effectively local in the $w$ direction when ZSMs are absent, underscoring the necessity of ZSMs in nonlocal braiding statistics along $w$.

\subsection{General theory of Toeplitz braiding and ZSMs}

Given a more general example, how can we understand the correspondence between ZSMs and nonlocal braiding statistics along the stacking direction in a more general framework? 
For more general i$BF$ theories with Toeplitz $K$ matrices, the connection between nonlocal braiding statistics and ZSMs can also be understood from the SVD of the $K$ matrix. 
The SVD of an arbitrary $\  N\times  N$ matrix $ K$ (not necessarily Toeplitz) is written as
 $  K=\sum_{i=1}^{ N}\sigma_i\mathbf u_i \mathbf v_i^\dag=\sigma_1\mathbf u_1 \mathbf v_1^\dag+\sigma_2\mathbf u_2 \mathbf v_2^\dag+\cdots$, where the left singular modes $\{\mathbf u_i\}$, right singular modes $\{\mathbf v_i\}$ and the corresponding singular values $\{\sigma_i\}$ of  $K$ are determined by 
 $
     K \mathbf v_i=\sigma_i \mathbf u_i,~ K^\dag \mathbf u_i=\sigma_i \mathbf v_i,\quad i=1,\ldots, N$ \cite{brunelli2023restoration,golub2013matrix}.
All singular modes can be chosen to be mutually orthonormal, i.e.,
$\mathbf u_i^\dagger \mathbf u_j = \delta_{i,j}, ~ \mathbf v_i^\dagger \mathbf v_j = \delta_{i,j}$. 
The singular values are nonnegative by definition, and for any real matrix the corresponding singular modes can be chosen to be real\footnote{As the $K$ matrices in this work are real matrices, we use the transpose ($\mT$) rather than the Hermitian conjugate ($\dag$) in all expressions related to the SVD of Toeplitz $K$ matrices hereafter.}.
Moreover, the LZSMs, RZSMs and the singular values in this paper may differ from the conventional definitions by a sign, which can be migrated among them without affecting the corresponding terms in SVD.

 In the following, LZSMs 
  are denoted by $\mathbf u_1,~\mathbf u_2,\cdots$ and the RZSMs
    are denoted by $\mathbf v_1,~\mathbf v_2,\cdots$.
For finite system size $N$, we write
 $	K = \sum_i \sigma_i \, \mathbf{u}_i \mathbf{v}_i^\mT$, 
so its inverse is given by
$	K^{-1} = \sum_i \sigma_i^{-1} \, \mathbf{v}_i \mathbf{u}_i^\mT$.
If the $K$ matrix possesses ZSMs, and all the LZSMs and RZSMs, namely $\mathbf{u}_j$ and $\mathbf{v}_j$, are  located at the opposite $w$ boundaries, then the ZSMs and their exponentially small singular values dominate the contribution to upper-right or lower-left elements of $K^{-1}$.

More specifically, if all the LZSMs localize at $w_1$ boundary and all the RZSMs localize at $w_2$ boundary, then, the ZSMs and their exponentially small singular values dominate the contribution to lower-left elements of $K^{-1}$ [see Fig.~\ref{particleloopbraiding}(e), column 1];
if all the LZSMs localize at $w_2$ boundary and all the RZSMs localize at $w_1$ boundary, then, the ZSMs and their exponentially small singular values dominate the contribution to upper-right elements of $K^{-1}$ [see Fig.~\ref{particleloopbraiding}(e), column 2].  
In both cases, the bulk modes contribute exponentially small values to lower-left or upper-right elements of $K^{-1}$.
 That is,
\begin{gather}
	(K^{-1})_{I,J} \approx \left( \sum_{j \in \text{ZSMs}} \frac{1}{\sigma_j} \, \mathbf{v}_j \mathbf{u}_j^\mT\right)_{I,J}\label{approxgeneral}
\end{gather}
for $I\sim 1$, $J\sim Nd$ or $I\sim Nd$, $J\sim 1$, where $d$ is the size of the block $A$, block $B$ and block $C$ in Eq.~(\ref{kmatrix}). Therefore, the corresponding particle-loop braiding phase $\Theta_{I,J}=2\pi (K^{-1})_{I,J}$, is also dominated by the contribution of ZSMs. 

In particular, certain $K$ matrices can support multiple LZSMs localized at both $w$ boundaries, with RZSMs exhibiting similar pattern [see Fig.~\ref{particleloopbraiding}(e), column 3]. In this case, nonzero elements exist for both the upper-right and lower-left regions of $K^{-1}$. These elements, together with the corresponding particle-loop braiding phases, are also dominated by the contributions from ZSMs. Concrete examples of this case are presented in Sec.~\ref{sectionIIIB}.

 It is noteworthy that the presence of ZSMs does not necessarily lead to nontrivial braiding statistics between particles and loops located at opposite $w$ boundaries, since ZSMs alone do not guarantee the existence of noninteger elements in the upper-right or lower-left elements of $K^{-1}$, which are required for nontrivial braiding statistics.
If the upper-right and lower-left elements of $K^{-1}$ are all integers, the corresponding braiding phases $2\pi (K^{-1})_{I,J}$  \textit{trivially} take values of  multiples of $2\pi$, which indicates that the nontrivial particle-loop braiding phases vanish for particles and loops placed sufficiently far away in the $w$ direction. For example, if $n\, \rad (n)\divides b$, then the $(K_{\mathrm{HN}}^{-1})_{I,J}$ in Eq.~(\ref{hninverseele}) is integer for sufficiently large $|I-J|$. More concretely, $K_{\mathrm{HN}}$ with $n=2$, $b=4$ is an example where ZSMs are present but nonlocal braiding statistics along $w$ direction is absent. 
The braiding phase $\Theta_{I,J} = \pi\,\delta_{I,J}$ is nonzero only for $I=J$, indicating that nontrivial braiding statistics is nontrivial only for particles and loops residing in the same layer.
Nevertheless, such special cases are very dilute in the whole parameter space; hence, the appearance of ZSMs indeed indicates nonlocal braiding statistics along the $w$ direction in most cases.

The same mathematical structure emerges in non-Hermitian directional amplification studies, which stem from researches in driven-dissipative systems \cite{porras2019topological,PhysRevA.103.033513,brunelli2023restoration,PhysRevB.103.L241408}. 
One-dimensional driven-dissipative cavity arrays can display directional amplification of coherent inputs, with gains detected at the other boundary scale exponentially with system size. 
When the input is switched to the other end, the amplification effect is suppressed.
 Moreover, such exotic amplification effect absent in Hermitian systems can also be naturally understood from the singular value decomposition of the non-Hermitian Hamiltonian $H$.
   In systems with non-Hermitian directional amplification,
  the SVD spectrum of $H$ under OBC exhibits ZSMs, whose corresponding LZSMs $\{\mathbf u_j\}$ and RZSMs $\{\mathbf v_j\}$, localize exponentially at opposite edges. In the linear-response regime, the susceptibility \(\chi = H^{-1} = \sum_j \sigma_j^{-1} \mathbf v_j \mathbf u_j^\dagger\) is thus dominated by contributions from these ZSMs, leading to exponentially large, unidirectional transmission between boundaries \cite{brunelli2023restoration}. This SVD-based framework provides a mathematically transparent diagnosis
    to directional amplification of physical signals in driven-dissipative systems.

\begin{figure*}[htbp]
    \centering
    \includegraphics[width=.85\textwidth]{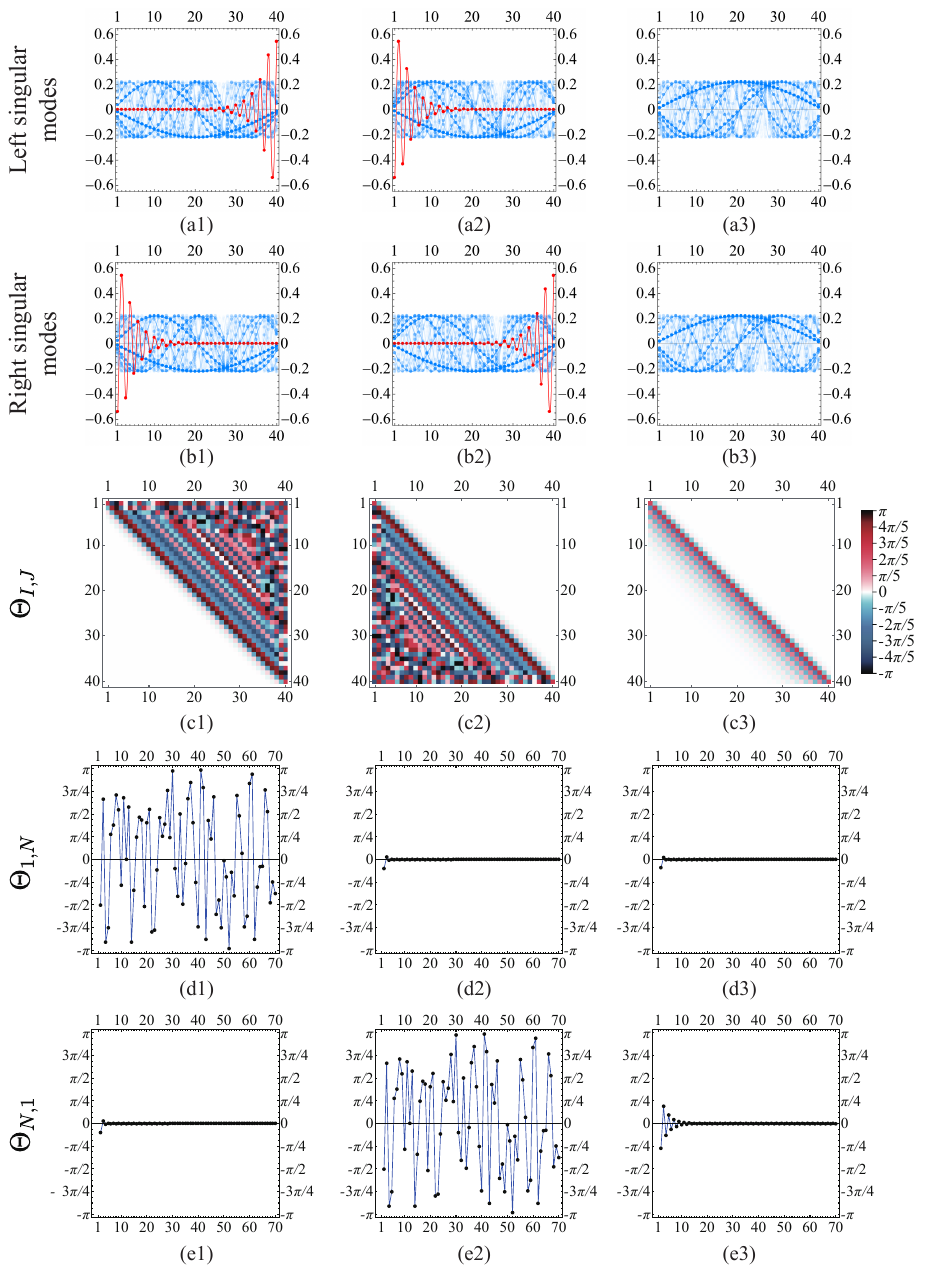}     \caption{
 Braiding statistics encoded in i$BF$ theories with Hatano-Nelson-type (HN-type) $K$ matrices. Panels (a1)--(e1) and (a2)--(e2) demonstrate the Toeplitz braiding encoded in HN-type i$BF$ theories, with the $K_{\mathrm{HN}}'$ matrix specified by the parameters $n=5$, $b=1$, $c=5$ and $n=5$, $b=5$, $c=1$, respectively. Panels (a3)--(e3) demonstrate a trivial i$BF$ theory with $K_{\mathrm{HN}}'$ matrix specified by $n=5$, $b=3$ and $c=1$. Panels (a1)--(a3) are the plots of the left singular modes of these matrices, while panels (b1)--(b3) are the plots of the right singular modes of these matrices. 
   Panels (c1)--(c3)  are the matrix plots of the braiding phases $\Theta_{I,J}=2\pi ({K_{\mathrm{HN}}'}^{-1})_{I,J}$. 
     For illustrative purpose, we take the system size $N$ to be 40. Panels (d1)--(d3) and (e1)--(e3) demonstrate how the braiding phases $\Theta_{1,N}$ and $\Theta_{N,1}$ vary as the system size $N$ increases.}
    \label{hnbraiding}
\end{figure*}

\subsection{Numerical results}

In Fig.~\ref{hnbraiding}, we demonstrate the Toeplitz braiding encoded in i$BF$ theories with $K$ matrices sharing the same mathematical form of more general Hatano-Nelson Hamiltonians [Eq.~(\ref{kmatrixhatano})], verifying the results obtained from analytic deduction.

Figures \ref{hnbraiding}(a1)--\ref{hnbraiding}(e1) demonstrate the braiding statistics encoded in HN-type i$BF$ theories with the $K_{\mathrm{HN}}'$ matrix specified by $n=5$, $b=1$, $c=5$. Figures \ref{hnbraiding}(a1) and \ref{hnbraiding}(b1) are the plots of the singular modes of the corresponding $K_{\mathrm{HN}}'$ matrix, and ZSMs are highlighted in red. Figure \ref{hnbraiding}(c1) is the matrix plot of the braiding phase $\Theta_{I,J}=2\pi ({K_{\mathrm{HN}}'}^{-1})_{I,J}$ introduced in Eq.~(\ref{thetaij}).
The presence of nontrivial braiding phases in the upper-right region of ${K_{\mathrm{HN}}'}^{-1}$ indicates that there exist particle excitations located at $w_1$ boundary and loop excitations located at $w_2$ boundary, which can feel each other via braiding process. In contrast, the absence of the lower-left components of ${K_{\mathrm{HN}}'}^{-1}$ shows that the   braiding phase vanishes when particle's and loop's $w$ coordinates are exchanged.   As the number of layers $N$ goes to infinity, the braiding phase $\Theta_{1,N}$ in Fig.~\ref{hnbraiding}(d1) oscillates between $(-\pi,\pi]$, while   $\Theta_{N,1}$ in Fig.~\ref{hnbraiding}(e1) rapidly decays  to zero.  

\begin{figure}[htbp]
    \centering
    \includegraphics[width=\linewidth]{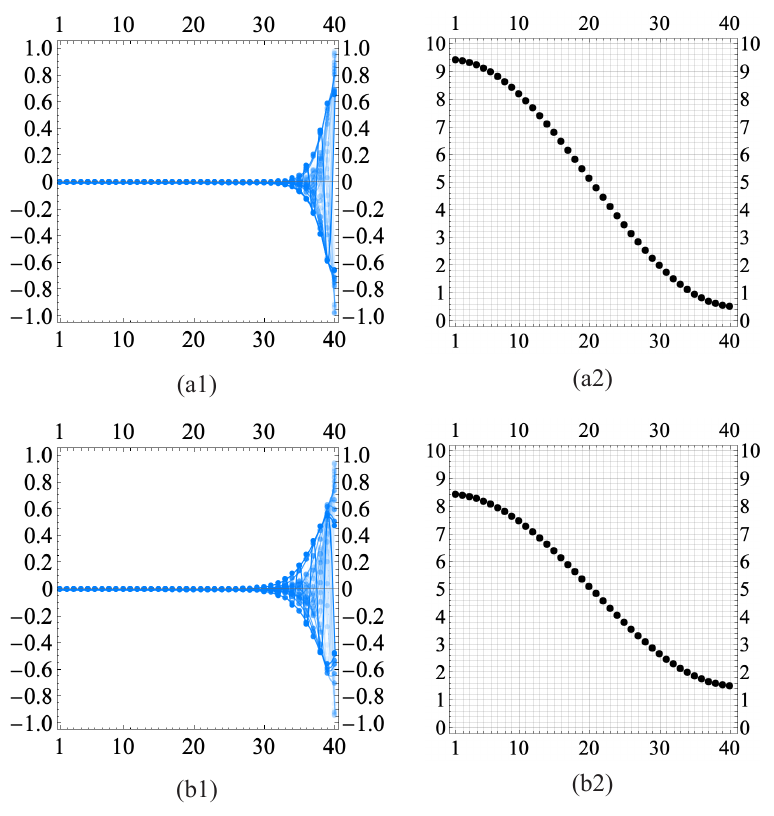}
    \caption{Panels (a1) and (a2) display skin modes and eigenvalues of $K_{\mathrm{HN}}'$ matrix specified by $n=5,b=5,c=1$. Panels (b1) and (b2) display skin modes and eigenvalues of $K_{\mathrm{HN}}'$ matrix specified by $n=5,b=3,c=1$. For illustrative purpose, we depict the skin modes and eigenvalues of the $K_{\mathrm{HN}}'$ matrices with system size $N=40$.}
    \label{compare}
\end{figure}

Likewise, Figs.~\ref{hnbraiding}(a2)--\ref{hnbraiding}(e2) demonstrate the braiding statistics encoded in NH-type i$BF$ theories with the $K_{\mathrm{HN}}'$ matrix specified by $n=5$, $b=5$, $c=1$, in which particle excitations located at $w_2$ boundary and loop excitations located at $w_1$ boundary that can feel each other via braiding process. Figures \ref{hnbraiding}(a2) and \ref{hnbraiding}(b2) are the plots of the singular modes of the $K_{\mathrm{HN}}'$ matrix, and ZSMs are highlighted in red.
Figure \ref{hnbraiding}(c2) is the matrix plot of the braiding phase $\Theta_{I,J}=2\pi ({K_{\mathrm{HN}}'}^{-1})_{I,J}$ introduced in Eq.~(\ref{thetaij}).
The presence of nontrivial braiding phases in the lower-left region of ${K_{\mathrm{HN}}'}^{-1}$ indicates that there exist particle excitations located at $w_2$ boundary and loop excitations located at $w_1$ boundary, which can feel each other via braiding process. In contrast, the absence of the upper-right components of ${K_{\mathrm{HN}}'}^{-1}$ shows that the   braiding phase vanishes when particle's and loop's $w$ coordinates are exchanged. As the number of layers $N$ goes to infinity, the braiding phase $\Theta_{N,1}$ in Fig.~\ref{hnbraiding}(e2) oscillates between $(-\pi,\pi]$, while   $\Theta_{1,N}$ in Fig.~\ref{hnbraiding}(d2) rapidly decays  to zero.

For comparison, we also present an i$BF$ theory without Toeplitz braiding in Figs.~\ref{hnbraiding}(a3)--\ref{hnbraiding}(e3). The corresponding $K_{\mathrm{HN}}'$ matrix is obtained by the parameters $n=5$, $b=3$, and $c=1$. Figures \ref{hnbraiding}(a3) and \ref{hnbraiding}(b3) display the singular modes of the $K_{\mathrm{HN}}'$ matrix, where ZSMs are absent. The exponentially suppressed upper-right and lower-left elements in Fig.~\ref{hnbraiding}(c3) indicate the absence of nonlocal braiding statistics along the stacking direction, which is further reflected in the asymptotic convergence of $\Theta_{1,N}$ and $\Theta_{N,1}$, as shown in Figs.~\ref{hnbraiding}(d3) and \ref{hnbraiding}(e3).

To furthermore highlight the importance and necessity of  SVD in analyzing Toeplitz braiding encoded in asymmetric $K$ matrices, the skin modes and eigenvalues of the $K_{\mathrm{HN}}'$ matrices are shown in Fig.~\ref{compare}. Figures~\ref{compare}(a1) and \ref{compare}(a2) display the skin modes~\cite{MartinezNonHermitian2018,yaoEdge2018,yokomizoNonBloch2019,SebastianTopolgical2020,XiaononHermitian2020,helbigGeneralized2020,zhangCorrespondence2020,ZhangObervation2021,LiangDynamic2022,zhangUniversal2022,ZhaoTwo2025,HetenyiLocalized2025} and eigenvalues of the $K_{\mathrm{HN}}'$ matrix with parameters $n=5$, $b=5$, and $c=1$, whose corresponding braiding statistics is illustrated in Figs.~\ref{hnbraiding}(a2)--(e2). Similarly, Figs.~\ref{compare}(b1) and \ref{compare}(b2) show the skin modes and eigenvalues of the $K_{\mathrm{HN}}'$ matrix with parameters $n=5$, $b=3$, and $c=1$, whose braiding statistics is illustrated in Figs.~\ref{hnbraiding}(a3)--(e3). As seen from these comparisons, the skin modes and eigenvalues of the $K$ matrices of i$BF$ theories with and without Toeplitz braiding exhibit qualitatively similar features. Therefore, adopting the singular value decomposition perspective of the $K$ matrices is essential for a complete characterization.

\section{i$BF$ theory with non-Hermitian Su-Schrieffer-Heeger type $K$ matrix\label{sectionIIIB}}

\subsection{Toeplitz Braiding encoded in Su-Schrieffer-Heeger type $K$ matrix}
To gain further insight into Toeplitz braiding encoded in asymmetric $K$ matrices, we now turn to another analytically tractable example, which has richer phase structure as the parameters in $K$ change.  The $K$ matrix of such i$BF$ theory is
\begin{gather}
K_{\mathrm{nSSH}} = \left( \begin{matrix}
    A & C & & & & \\
    B & A & C & & & \\
    & B & A & C & & \\
    & & \ddots & \ddots & \ddots & \\
    & & & B & A & C \\
    & & & & B & A \\
    \end{matrix} \right),\label{nsshkmatrix}
\end{gather}
where the blocks $A$, $B$ and $C$ read
\begin{gather}
	A=\begin{pmatrix}
		0 & n_1 \\ n_2 & 0
	\end{pmatrix},\quad B=\begin{pmatrix}
		0 & m_1 \\ 0 & 0
	\end{pmatrix},\quad C= \begin{pmatrix}
		0 & 0 \\ m_2 & 0
	\end{pmatrix}.
\end{gather}
The $K_{\mathrm{nSSH}}$ matrix exhibits the same mathematical structure as that of the non-Hermitian SSH (nSSH) model \cite{PhysRevA.89.062102,PhysRevB.97.045106}. By applying Cramer's rule, one readily unveils the inverse of $K_{\mathrm{nSSH}}$:
\begin{gather}
	(K_{\mathrm{nSSH}}^{-1})_{I,J}=\notag \\
	\left\{\begin{array}{l}
		\frac{1}{n_1}\left(-\frac{m_1}{n_1}\right)^{\frac{I-J-1}{2}},\quad I>J, I\in 2\mathbb Z^+,J\in 2\mathbb Z^+ -1 ;\\
		\frac{1}{n_2}\left(-\frac{m_2}{n_2}\right)^{\frac{J-I-1}{2}},\quad I<J, I\in 2\mathbb Z^+-1,J\in 2\mathbb Z^+  ;\\
		0,\quad \text{others}.
	\end{array}\right.
\end{gather}

For instance, when the system size $N=4$, $K_{\mathrm{nSSH}}^{-1}$ reads
\begin{gather}
    \left(
\begin{array}{cccccccc}
 0 & \frac{1}{n_2} & 0 & -\frac{m_2}{n_2^2} & 0 & \frac{m_2^2}{n_2^3} & 0 & -\frac{m_2^3}{n_2^4} \\
 \frac{1}{n_1} & 0 & 0 & 0 & 0 & 0 & 0 & 0 \\
 0 & 0 & 0 & \frac{1}{n_2} & 0 & -\frac{m_2}{n_2^2} & 0 & \frac{m_2^2}{n_2^3} \\
 -\frac{m_1}{n_1^2} & 0 & \frac{1}{n_1} & 0 & 0 & 0 & 0 & 0 \\
 0 & 0 & 0 & 0 & 0 & \frac{1}{n_2} & 0 & -\frac{m_2}{n_2^2} \\
 \frac{m_1^2}{n_1^3} & 0 & -\frac{m_1}{n_1^2} & 0 & \frac{1}{n_1} & 0 & 0 & 0 \\
 0 & 0 & 0 & 0 & 0 & 0 & 0 & \frac{1}{n_2} \\
 -\frac{m_1^3}{n_1^4} & 0 & \frac{m_1^2}{n_1^3} & 0 & -\frac{m_1}{n_1^2} & 0 & \frac{1}{n_1} & 0 \\
\end{array}
\right).
\end{gather}

The classification of $K_{\mathrm{nSSH}}^{-1}$ can be described across four parameter regimes, each determined by the relative magnitudes of $|m_1|$ and $|n_1|$ as well as $|m_2|$ and $|n_2|$. These parameter regimes, combined with the divisibility conditions described below, determine the presence or absence of Toeplitz braiding along the stacking direction.
 
\textbf{Case I} corresponds to $|m_1| > |n_1|$ and $|m_2| < |n_2|$, where the upper-right elements of $K_{\mathrm{nSSH}}^{-1}$ are suppressed and the lower-left elements increase exponentially with system size. In this regime, nontrivial braiding statistics occurs between particles at the $w_2$ boundary and loops at the $w_1$ boundary when $n_1\,\mathrm{rad}(n_1)\nmid m_1$; otherwise, when $n_1\,\mathrm{rad}(n_1) | m_1$, the braiding statistics between particles and loops at distinct $w$ boundaries is trivial. 

\textbf{Case II} arises when $|m_1| < |n_1|$ and $|m_2| > |n_2|$, leading to $K_{\mathrm{nSSH}}^{-1}$ with exponentially suppressed lower-left elements, while the upper-right elements increase exponentially with the system size. Here, particles at the $w_1$ boundary can braid nontrivially with loops at the $w_2$ boundary if $n_2\,\mathrm{rad}(n_2)\nmid m_2$, while exchanging the $w$ coordinates of the particles and the loops yields trivial statistics. If $n_2\,\mathrm{rad}(n_2) | m_2$, then the braiding statistics between particles and loops at distinct $w$ boundaries is trivial. 

\textbf{Case III} occurs when $|m_1| > |n_1|$ and $|m_2| > |n_2|$, so both nonzero upper-right and lower-left elements exist. In this regime, several possibilities arise: nontrivial braiding between $w_2$-boundary particles and $w_1$-boundary loops when $n_1\,\mathrm{rad}(n_1)\nmid m_1$ but $n_2\,\mathrm{rad}(n_2)\mid m_2$; nontrivial braiding between $w_1$-boundary particles and $w_2$-boundary loops when the conditions are reversed, i.e. $n_2\,\mathrm{rad}(n_2)\nmid m_2$ but $n_1\,\mathrm{rad}(n_1)\mid m_1$; nontrivial braiding statistics for both configurations of particles and loops located at distinct $w$ boundaries when neither divisibility condition holds; fully trivial statistics between particles and loops at distinct $w$ boundaries when both divisibility conditions hold.

Finally, \textbf{Case IV}, defined by $|m_1| < |n_1|$ and $|m_2| < |n_2|$, has both upper-right and lower-left elements of $K_{\mathrm{nSSH}}^{-1}$ suppressed and yields only trivial braiding statistics between particles and loops located at distinct $w$ boundaries.

\subsection{ZSMs and Toeplitz braiding}

To deepen our understanding of the origin of this nonlocal braiding statistics along the $w$ direction, i.e. Toeplitz braiding, the relation between ZSMs of the $K$ matrices and Toeplitz braiding  in i$BF$ theories with nSSH-type $K$ matrix is further investigated.
In Appendix \ref{detailednssh}, the ZSMs are derived analytically. In Case I where $\left\{\begin{array}{l}
	|m_1| > |n_1|\\|m_2| < |n_2|
\end{array}\right.$, $K_{\mathrm{nSSH}}$ possesses the following LZSM and RZSM, denoted by $\mathbf u_1$ and $\mathbf v_1$, respectively.
\begin{gather}
	\mathbf u_1=\sqrt{\frac{1-\frac{n_1^2}{m_1^2}}{1-(\frac{n_1}{m_1})^{2N}}}\begin{pmatrix}
		 1 & 0 & -\frac{n_1}{m_1} & 0 &\cdots & \left(-\frac{n_1}{m_1}\right)^{N-1} & 0
	\end{pmatrix}^\mT,\\
	\mathbf v_1=\sqrt{\frac{1-\frac{m_1^2}{n_1^2}}{1-(\frac{m_1}{n_1})^{2N}}}\begin{pmatrix}
		0 & 1 & 0 & -\frac{m_1}{n_1} & \cdots & 0 & \left(-\frac{m_1}{n_1}\right)^{N-1}
	\end{pmatrix}^\mT.
\end{gather}
In Case II where $\left\{\begin{array}{l}
	|m_1| < |n_1|\\|m_2| > |n_2|
\end{array}\right.$, $K_{\mathrm{nSSH}}$ possesses the following LZSM and RZSM, denoted by $\mathbf u_2$ and $\mathbf v_2$, respectively,
\begin{gather}
	\mathbf u_2=\sqrt{\frac{1-\frac{m_2^2}{n_2^2}}{1-(\frac{m_2}{n_2})^{2N}}}\begin{pmatrix}
		0 & 1 & 0 & -\frac{m_2}{n_2} & \cdots & 0 & \left(-\frac{m_2}{n_2}\right)^{N-1}
	\end{pmatrix}^\mT,\\
	\mathbf v_2=\sqrt{\frac{1-\frac{n_2^2}{m_2^2}}{1-(\frac{n_2}{m_2})^{2N}}}\begin{pmatrix}
		 1 & 0 & -\frac{n_2}{m_2} & 0 &\cdots & \left(-\frac{n_2}{m_2}\right)^{N-1} & 0
	\end{pmatrix}^\mT.
\end{gather}
In Case III where $\left\{\begin{array}{l}
	|m_1| > |n_1|\\|m_2| > |n_2|
\end{array}\right.$, the $K_{\mathrm{nSSH}}$ matrix possesses two sets of ZSMs, $\mathbf u_1$, $\mathbf v_1$ and $\mathbf u_2$, $\mathbf v_2$. In Case IV where $\left\{\begin{array}{l}
	|m_1| < |n_1|\\|m_2| < |n_2|
\end{array}\right.$, the $K_{\mathrm{nSSH}}$ matrix possesses no ZSM.
Furthermore, there is an exponentially small violation from the decaying tails of ZSMs to the opposite boundary, leading to exponentially small singular values $\sigma_1$ and $\sigma_2$, which are obtained from
\begin{gather}
	\sigma_1=\mathbf u_1^\mT K_{\mathrm{nSSH}} \mathbf v_1=\sqrt{\frac{1-\frac{m_1^2}{n_1^2}}{1-(\frac{m_1}{n_1})^{2N}}} \sqrt{\frac{1-\frac{n_1^2}{m_1^2}}{1-(\frac{n_1}{m_1})^{2N}}} n_1,\\
	\sigma_2=\mathbf u_2^\mT K_{\mathrm{nSSH}} \mathbf v_2=\sqrt{\frac{1-\frac{n_2^2}{m_2^2}}{1-(\frac{n_2}{m_2})^{2N}}}\sqrt{\frac{1-\frac{m_2^2}{n_2^2}}{1-(\frac{m_2}{n_2})^{2N}}} n_2.
\end{gather}
 In line with Eq.~(\ref{approxgeneral}), we construct an approximate matrix $M_{\mathrm{nSSH}}$ for the inverse $K_{\mathrm{nSSH}}^{-1}$ that captures the contribution of ZSMs to the upper-right or lower-left elements of $K_{\mathrm{nSSH}}^{-1}$. For Case I with $\left\{\begin{array}{l}
     |m_1| > |n_1|  \\
     |m_2| < |n_2| 
\end{array}\right.$, $M_{\mathrm{nSSH}}$ is constructed by $\mathbf u_1$, $\mathbf v_1$ and $\sigma_1$:
\begin{gather}
	M_{\mathrm{nSSH}}=\frac{1}{\sigma_1}\mathbf v_1 \mathbf u_1^\mT,\quad (M_{\mathrm{nSSH}})_{I,J}=\notag \\
	\left\{ \begin{array}{l}
		\frac{1}{n_1}\left(-\frac{m_1}{n_1}\right)^{(I-J-1)/2},\quad I>J,\ I\in 2\ZZ^+,\ J\in 2\ZZ^+-1;\\
		-\frac{1}{m_1}\left(-\frac{n_1}{m_1}\right)^{(J-I-1)/2},\quad J>I,\ I\in 2\ZZ^+,\ J\in 2\ZZ^+-1;\\
		0,\quad \mathrm{others}.
	\end{array}\right. 
\end{gather}
For example, if the system size $N=4$, then
\begin{gather}
	M_{\mathrm{nSSH}}=\left(
\begin{array}{cccccccc}
 0 & 0 & 0 & 0 & 0 & 0 & 0 & 0 \\
 \frac{1}{n_1} & 0 & -\frac{1}{m_1} & 0 & \frac{n_1}{m_1^2} & 0 & -\frac{n_1^2}{m_1^3} & 0 \\
 0 & 0 & 0 & 0 & 0 & 0 & 0 & 0 \\
 -\frac{m_1}{n_1^2} & 0 & \frac{1}{n_1} & 0 & -\frac{1}{m_1} & 0 & \frac{n_1}{m_1^2} & 0 \\
 0 & 0 & 0 & 0 & 0 & 0 & 0 & 0 \\
 \frac{m_1^2}{n_1^3} & 0 & -\frac{m_1}{n_1^2} & 0 & \frac{1}{n_1} & 0 & -\frac{1}{m_1} & 0 \\
 0 & 0 & 0 & 0 & 0 & 0 & 0 & 0 \\
 -\frac{m_1^3}{n_1^4} & 0 & \frac{m_1^2}{n_1^3} & 0 & -\frac{m_1}{n_1^2} & 0 & \frac{1}{n_1} & 0 \\
\end{array}
\right).
\end{gather}

For Case II with $\left\{\begin{array}{l}
     |m_1| < |n_1|  \\
     |m_2| > |n_2| 
\end{array}\right.$, the approximate matrix $M_{\mathrm{nSSH}}$ is constructed by $\mathbf u_2$, $\mathbf v_2$ and $\sigma_2$.
\begin{gather}
	M_{\mathrm{nSSH}}=\frac{1}{\sigma_2} \mathbf v_2 \mathbf u_2^\mT,\quad (M_{\mathrm{nSSH}})_{I,J}= \notag \\
	\left\{ \begin{array}{l}
		\frac{1}{n_2}\left(-\frac{m_2}{n_2}\right)^{(J-I-1)/2},\quad J>I,\ I\in 2\ZZ^+-1,\ J\in 2\ZZ^+;\\
		-\frac{1}{m_2}\left(-\frac{n_2}{m_2}\right)^{(I-J-1)/2},\quad I>J,\ I\in 2\ZZ^+-1,\ J\in 2\ZZ^+;\\
		0,\quad \mathrm{others}.
	\end{array}\right. 
\end{gather}
For example, if system size $N=4$, then
\begin{gather}
	M_{\mathrm{nSSH}}=\left(
\begin{array}{cccccccc}
 0 & \frac{1}{n_2} & 0 & -\frac{m_2}{n_2^2} & 0 & \frac{m_2^2}{n_2^3} & 0 & -\frac{m_2^3}{n_2^4} \\
 0 & 0 & 0 & 0 & 0 & 0 & 0 & 0 \\
 0 & -\frac{1}{m_2} & 0 & \frac{1}{n_2} & 0 & -\frac{m_2}{n_2^2} & 0 & \frac{m_2^2}{n_2^3} \\
 0 & 0 & 0 & 0 & 0 & 0 & 0 & 0 \\
 0 & \frac{n_2}{m_2^2} & 0 & -\frac{1}{m_2} & 0 & \frac{1}{n_2} & 0 & -\frac{m_2}{n_2^2} \\
 0 & 0 & 0 & 0 & 0 & 0 & 0 & 0 \\
 0 & -\frac{n_2^2}{m_2^3} & 0 & \frac{n_2}{m_2^2} & 0 & -\frac{1}{m_2} & 0 & \frac{1}{n_2} \\
 0 & 0 & 0 & 0 & 0 & 0 & 0 & 0 \\
\end{array}
\right).
\end{gather}

For Case III with $\left\{\begin{array}{l}
     |m_1| > |n_1|  \\
     |m_2| > |n_2| 
\end{array}\right.$, the approximate matrix $M_{\mathrm{nSSH}}$ is constructed by all $\mathbf u_i$, $\mathbf v_i$ and $\sigma_i$ ($i=1,2$).
\begin{gather}
	M_{\mathrm{nSSH}}=\frac{1}{\sigma_1}\mathbf v_1 \mathbf u_1^\mT+\frac{1}{\sigma_2} \mathbf v_2 \mathbf u_2^\mT,\quad (M_{\mathrm{nSSH}})_{I,J}= \notag \\
	\left\{ \begin{array}{l}
	\frac{1}{n_1}\left(-\frac{m_1}{n_1}\right)^{(I-J-1)/2},\quad I>J,\ I\in 2\ZZ^+,\ J\in 2\ZZ^+-1;\\
	-\frac{1}{m_1}\left(-\frac{n_1}{m_1}\right)^{(J-I-1)/2},\quad J>I,\ I\in 2\ZZ^+,\ J\in 2\ZZ^+-1;\\
		\frac{1}{n_2}\left(-\frac{m_2}{n_2}\right)^{(J-I-1)/2},\quad J>I,\ I\in 2\ZZ^+-1,\ J\in 2\ZZ^+;\\
		-\frac{1}{m_2}\left(-\frac{n_2}{m_2}\right)^{(I-J-1)/2},\quad I>J,\ I\in 2\ZZ^+-1,\ J\in 2\ZZ^+;\\
		0,\quad \mathrm{others}.
	\end{array}\right.\label{caseiiim2}
\end{gather}
For example, if system size $N=4$, then
\begin{gather}
 M_{\mathrm{nSSH}}=\notag \\ \left(
\begin{array}{cccccccc}
 0 & \frac{1}{n_2} & 0 & -\frac{m_2}{n_2^2} & 0 & \frac{m_2^2}{n_2^3} & 0 & -\frac{m_2^3}{n_2^4} \\
 \frac{1}{n_1} & 0 & -\frac{1}{m_1} & 0 & \frac{n_1}{m_1^2} & 0 & -\frac{n_1^2}{m_1^3} & 0 \\
 0 & -\frac{1}{m_2} & 0 & \frac{1}{n_2} & 0 & -\frac{m_2}{n_2^2} & 0 & \frac{m_2^2}{n_2^3} \\
 -\frac{m_1}{n_1^2} & 0 & \frac{1}{n_1} & 0 & -\frac{1}{m_1} & 0 & \frac{n_1}{m_1^2} & 0 \\
 0 & \frac{n_2}{m_2^2} & 0 & -\frac{1}{m_2} & 0 & \frac{1}{n_2} & 0 & -\frac{m_2}{n_2^2} \\
 \frac{m_1^2}{n_1^3} & 0 & -\frac{m_1}{n_1^2} & 0 & \frac{1}{n_1} & 0 & -\frac{1}{m_1} & 0 \\
 0 & -\frac{n_2^2}{m_2^3} & 0 & \frac{n_2}{m_2^2} & 0 & -\frac{1}{m_2} & 0 & \frac{1}{n_2} \\
 -\frac{m_1^3}{n_1^4} & 0 & \frac{m_1^2}{n_1^3} & 0 & -\frac{m_1}{n_1^2} & 0 & \frac{1}{n_1} & 0 \\
\end{array}
\right).
\end{gather}

For all the cases with Toeplitz braiding, in the thermodynamic limit $N\rightarrow\infty$, $M_{\mathrm{nSSH}}$ approximates the lower-left elements of $K_{\mathrm{nSSH}}^{-1}$ up to exponentially suppressed terms when $|m_1|>|n_1|$, and approximates the upper-right elements of $K_{\mathrm{nSSH}}^{-1}$ up to exponentially suppressed terms when $|m_2|>|n_2|$. Therefore, ZSMs together with the corresponding exponentially small singular values, capture the braiding statistics between particles and loops residing at distinct $w$ boundaries.
In particular, Case III demonstrates that nontrivial Toeplitz braiding between particles and loops located at both $w$ boundaries are possible, provided that two sets of ZSMs exist, and that both $w$ boundaries host localized LZSMs (RZSMs).

It should be noticed that the presence of ZSMs is not a sufficient condition for nontrivial Toeplitz braiding. For instance, even the parameters in $K_{\mathrm{nSSH}}$ fall in Case I, as long as $n_1 \,\mathrm{rad}(n_1) | m_1$, the lower-left elements of $K_{\mathrm{nSSH}}^{-1}$ are all integers, forbidding  nontrivial Toeplitz braiding along the $w$ direction. Nevertheless, such special cases are very dilute in the whole parameter space; hence, the appearance of ZSMs indeed indicates nonlocal braiding statistics along the $w$ direction in most cases.

\subsection{Numerical results}

\begin{figure*}
    \centering
    \includegraphics[width=.85\textwidth]{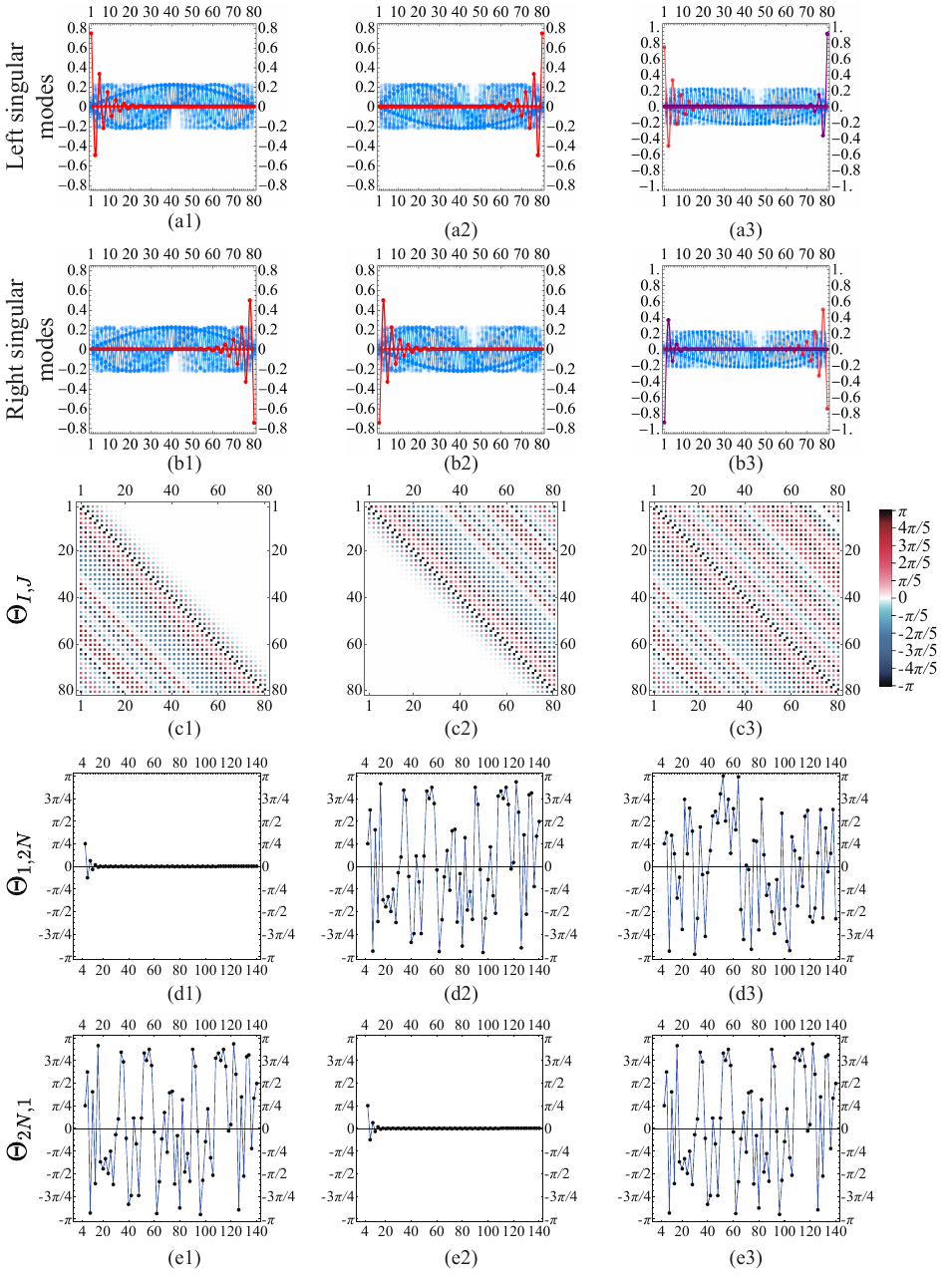}     \caption{Braiding statistics in different nSSH-type i$BF$ theories. Panels (a1)--(e1), (a2)--(e2) and (a3)--(e3) demonstrate the braiding statistics encoded in nSSH-type i$BF$ theories with the $K_{\mathrm{nSSH}}$ matrix specified by $\left\{\begin{array}{l}
    	n_1=n_2=2 \\
    	m_1=3, m_2=1
    \end{array}\right.$, $\left\{\begin{array}{l}
    	n_1=n_2=2 \\
    	m_1=1, m_2=3
    \end{array}\right.$ and $\left\{\begin{array}{l}
    	n_1=n_2=2 \\
    	m_1=3, m_2=5
    \end{array}\right.$, respectively. Panels (a1)--(a3)  are the plots of the left singular modes of these matrices, while panels (b1)--(b3) are the plots of the right singular modes of these matrices. The ZSMs are highlighted in red and purple.
   Panels (c1)--(c3)  are the matrix plots of the braiding phase $\Theta_{I,J}=2\pi (K_{\mathrm{nSSH}}^{-1})_{I,J}$, where we take the system size $N=40$ (the resulting matrix size of $K_{\mathrm{nSSH}}$ is $2N=80$) as illustrative examples.
   Panels (d1)--(d3) and (e1)--(e3) demonstrate how the braiding phases $\Theta_{1,2N}$ and $\Theta_{2N,1}$ vary as the system size $N$ increases.
     }
    \label{nsshbraiding}
\end{figure*}

In Fig.~\ref{nsshbraiding}, we depict several examples that demonstrate the nonlocal braiding statistics induced by ZSMs, verifying the analytic deductions presented above. Figures \ref{nsshbraiding}(a1)--\ref{nsshbraiding}(e1) demonstrates the braiding statistics encoded in nSSH-type i$BF$ theories with the $K_{\mathrm{nSSH}}$ matrix specified by $n_1=n_2=2$, $m_1=3$, $m_2=1$, which belongs to Case I.
 Figures \ref{nsshbraiding}(a1) and \ref{nsshbraiding}(b1) are the plots of the singular modes of the $K_{\mathrm{nSSH}}$ matrix, where ZSMs are highlighted in red. Figure \ref{nsshbraiding}(c1) is the matrix plot of the braiding phase $\Theta_{I,J}=2\pi (K_{\mathrm{nSSH}}^{-1})_{I,J}$ [Eq.~(\ref{thetaij})]. 
 The presence of nontrivial braiding phases in the lower-left region of $\Theta_{I,J}$ indicates that there exist particle excitations located at $w_2$ boundary and loop excitations located at $w_1$ boundary that can feel each other via braiding process. In contrast, the absence of a nontrivial braiding phase in the upper-right region of $\Theta_{I,J}$ shows that when their $w$ coordinates are exchanged, the braiding phase vanishes. Figure \ref{nsshbraiding}(d1) shows that as the size $2N$ of the matrix increases, the braiding phase $\Theta_{1,2N}$ drastically decays to zero as the system size increases, while Fig.~\ref{nsshbraiding}(e1) shows that the braiding phase $\Theta_{2N,1}$ oscillates between $(-\pi,\pi]$, reflecting the non-liquid nature of the i$BF$ theory.

Likewise, Figs.~\ref{nsshbraiding}(a2)--\ref{nsshbraiding}(e2) demonstrates the braiding statistics encoded in nSSH-type i$BF$ theories with the $K_{\mathrm{nSSH}}$ matrix specified by $n_1=n_2=2$, $m_1=1$, $m_2=3$, which belongs to Case II.
 Figures \ref{nsshbraiding}(a2) and \ref{nsshbraiding}(b2) are the plots of the singular modes of such $K_{\mathrm{nSSH}}$ matrix, where ZSMs are highlighted in red. Figure \ref{nsshbraiding}(c2) is the matrix plot of the braiding phase $\Theta_{I,J}=2\pi (K_{\mathrm{nSSH}}^{-1})_{I,J}$. The presence of nontrivial braiding phases in the upper-right region of $\Theta_{I,J}$ indicates that there exist particle excitations located at $w_1$ boundary and loop excitations located at $w_2$ boundary that can feel each other via braiding process. In contrast, the absence of a nontrivial braiding phase in the lower-left region of $\Theta_{I,J}$ shows that when their $w$ coordinates are exchanged, the braiding phase vanishes. Figure \ref{nsshbraiding}(d2) shows that as the system size $N$ increases, the braiding phase $\Theta_{1,2N}$ oscillates between $(-\pi,\pi]$, while FIg.~\ref{nsshbraiding}(e2) shows that the braiding phase $\Theta_{2N,1}$ drastically decays to zero as the system size $N$ increases.

 Figures \ref{nsshbraiding}(a3)--\ref{nsshbraiding}(e3) demonstrates the braiding statistics encoded in nSSH-type i$BF$ theories with the $K_{\mathrm{nSSH}}$ matrix specified by $n_1=n_2=2$, $m_1=3$, $m_2=5$, which belongs to Case III.
  In this case, the $K_{\mathrm{nSSH}}$ matrix possesses two sets of ZSMs, highlighted in red and purple in Figs.~\ref{nsshbraiding}(a3) and \ref{nsshbraiding}(b3). Nontrivial braiding statistics arises for both configurations: particles at $w_1$ boundary with loops at $w_2$ boundary, and vice versa, as is demonstrated in the matrix plot in Figs.~\ref{nsshbraiding}(c3) of the braiding phase $\Theta_{I,J}=2\pi (K_{\mathrm{nSSH}}^{-1})_{I,J}$. Figures \ref{nsshbraiding}(d3) and \ref{nsshbraiding}(e3) show that as the system size $N$ increases, both $\Theta_{1,2N}$ and $\Theta_{2N,1}$ oscillate between $(-\pi,\pi]$.

\subsection{Equivalence between eigendecomposition and SVD in capturing Toeplitz braiding for symmetric $K$ matrix\label{consistency}
}

So far, we focus on the role of ZSMs in nonlocal braiding statistics along the stacking direction.  Reference~\cite{li2024} shows that boundary zero modes obtained from eigendecompostion capture the upper-right and lower-left elements of $K^{-1}$ in the thermodynamic limit when the $K$ matrix is symmetric. Are these two approaches compatible with each other? 
Are there any differences between the approximate matrix $M$ constructed from ZSMs and the approximate matrix $M'$ constructed from boundary zero modes when the $K$ matrix is symmetric? 
In fact, Eqs.~(\ref{simplifiedsvdzero}) and (\ref{simplifiedzsm2}) in Appendix~\ref{appendix_detailed} are equations for the boundary zero modes  of the Hermitian $K$ matrices, thus these two approaches are indeed identical. We explicitly verify this using $K_{\mathrm{nSSH}}$ for the case $n_1 = n_2 = n$ and $m_1 = m_2 = m$.
 When $n_1 = n_2 = n$ and $m_1 = m_2 = m$, $K_{\mathrm{nSSH}}$ [Eq.~(\ref{nsshkmatrix})] is a symmetric matrix sharing the same mathematical form of the SSH Hamiltonian \cite{su1979solitons}.  If $|m|>|n|$, then the approximate matrix $M_{\mathrm{nSSH}}'$ for the inverse $K_{\mathrm{nSSH}}^{-1}$ constructed from boundary zero modes $\mathbf w_1$ and $\mathbf w_2$ reads \cite{li2024}
\begin{gather}
	M_{\mathrm{nSSH}}'=\frac{1}{\lambda_1}\mathbf{w}_1\mathbf{w}_1^\mT+\frac{1}{\lambda_{2}}\mathbf{w}_2\mathbf{w}_2^\mT,
\end{gather}
where
\begin{subequations}
	\begin{gather}
\begin{split}
     &\mathsmaller{\mathbf w_1=\frac{1}{\sqrt 2}\sqrt{\frac{1-\left(\frac{n}{m}\right)^2}{1-\left(\frac{n}{m}\right)^{2N}}}}\\ &\mathsmaller{\mathsmaller{\left(1 , \left(-\frac{n}{m}\right)^{N-1} , \left(-\frac{n}{m}\right) , \left(-\frac{n}{m}\right)^{N-2} , \left(-\frac{n}{m}\right)^2 , \cdots , \left(-\frac{n}{m}\right)^{N-1} , 1\right)^\mT}},
\end{split}\\
    \begin{split}
    & \mathsmaller{\mathbf w_2=\frac{1}{\sqrt 2}\sqrt{\frac{1-\left(\frac{n}{m}\right)^2}{1-\left(\frac{n}{m}\right)^{2N}}}}\\ &\mathsmaller{\mathsmaller{\left(-1 , \left(-\frac{n}{m}\right)^{N-1} , \frac{n}{m} , \left(-\frac{n}{m}\right)^{N-2} , \cdots , \left(-\frac{n}{m}\right) , -\left(-\frac{n}{m}\right)^{N-1} , 1\right)^\mT}}.
\end{split}
\end{gather}
\end{subequations}
and the exponentially small eigenvalues are
\begin{gather}
	\lambda_1=n\frac{1-\left(\frac{n}{m}\right)^2}{1-\left(\frac{n}{m}\right)^{2N}}\left(-\frac{n}{m}\right)^{N-1},\\  \lambda_2= -n\frac{1-\left(\frac{n}{m}\right)^2}{1-\left(\frac{n}{m}\right)^{2N}}\left(-\frac{n}{m}\right)^{N-1}.
\end{gather}
The approximate matrix constructed from ZSMs is given by Eq.~(\ref{caseiiim2}).
Comparing $M_{\mathrm{nSSH}}$ and $M_{\mathrm{nSSH}}'$, we discover that 
\begin{gather}
	(M_{\mathrm{nSSH}})_{I,J}=(M_{\mathrm{nSSH}}')_{I,J}=\notag \\
	\left\{ \begin{array}{l}
		\frac{1}{n}\left(-\frac{m}{n}\right)^{(I-J-1)/2},\quad I>J, I\in 2\ZZ^+,J\in 2\ZZ^+-1;\\
		-\frac{1}{m}\left(-\frac{n}{m}\right)^{(I-J-1)/2},\quad I>J, I\in 2\ZZ^+-1,J\in 2\ZZ^+;\\
		\frac{1}{n}\left(-\frac{m}{n}\right)^{(J-I-1)/2},\quad J>I, J\in 2\ZZ^+,I\in 2\ZZ^+-1;\\
		-\frac{1}{m}\left(-\frac{n}{m}\right)^{(J-I-1)/2},\quad J>I, J\in 2\ZZ^+-1,I\in 2\ZZ^+;\\
		0,\quad \mathrm{others}.
	\end{array} \right.
\end{gather}
 For example, if the system size is $N=4$, $M_{\mathrm{nSSH}}$, $M_{\mathrm{nSSH}}'$ are of size $2N=8$,
\begin{gather}
	M_{\mathrm{nSSH}}'=M_{\mathrm{nSSH}}=\notag \\ \left(
\begin{array}{cccccccc}
 0 & \frac{1}{n} & 0 & -\frac{m}{n^2} & 0 & \frac{m^2}{n^3} & 0 & -\frac{m^3}{n^4} \\
 \frac{1}{n} & 0 & -\frac{1}{m} & 0 & \frac{n}{m^2} & 0 & -\frac{n^2}{m^3} & 0 \\
 0 & -\frac{1}{m} & 0 & \frac{1}{n} & 0 & -\frac{m}{n^2} & 0 & \frac{m^2}{n^3} \\
 -\frac{m}{n^2} & 0 & \frac{1}{n} & 0 & -\frac{1}{m} & 0 & \frac{n}{m^2} & 0 \\
 0 & \frac{n}{m^2} & 0 & -\frac{1}{m} & 0 & \frac{1}{n} & 0 & -\frac{m}{n^2} \\
 \frac{m^2}{n^3} & 0 & -\frac{m}{n^2} & 0 & \frac{1}{n} & 0 & -\frac{1}{m} & 0 \\
 0 & -\frac{n^2}{m^3} & 0 & \frac{n}{m^2} & 0 & -\frac{1}{m} & 0 & \frac{1}{n} \\
 -\frac{m^3}{n^4} & 0 & \frac{m^2}{n^3} & 0 & -\frac{m}{n^2} & 0 & \frac{1}{n} & 0 \\
\end{array}
\right),
\end{gather}
demonstrating that these two approaches are indeed identical if $K_{\mathrm{nSSH}}$ is symmetric.

\section{Conclusion and outlook\label{conclusion}}

In this paper, we investigated four-dimensional fracton topological orders
within the framework of infinite-component $BF$ (i$BF$) theories and identified
boundary zero singular modes (ZSMs) as the fundamental mechanism underlying
nonlocal braiding statistics along the stacking direction, which we term
``Toeplitz braiding''. A defining feature of Toeplitz-braiding i$BF$ theories is
their extreme boundary sensitivity and directionality: nontrivial
particle--loop braiding phases arise only when the excitations occupy specific
opposite boundaries along the stacking direction, while exchanging their
boundary locations causes the braiding phase to vanish in the thermodynamic
limit. We demonstrated that, for asymmetric Toeplitz $K$ matrices, the ZSM
sector dominates the off-diagonal corner structure of $K^{-1}$ at large system
size, thereby encoding the essential braiding data of particles and loops
residing on opposite boundaries. Through combined analytical and numerical
studies of i$BF$ theories with Hatano--Nelson and non-Hermitian
Su--Schrieffer--Heeger-type $K$ matrices, we established a sharp and universal
correspondence between the existence and boundary localization of ZSMs and the
emergence of nontrivial Toeplitz braiding phases. These results elevate stacking
constructions of topological field theories from a descriptive tool to a
predictive framework for engineering exotic braiding phenomena in higher
dimensions. More broadly, identifying ZSMs as the operative degrees of freedom
behind Toeplitz braiding provides a unifying principle that connects fracton
physics and topological field theory with non-Hermitian physics, particularly
directional amplification phenomena.

Beyond condensed-matter realizations, we propose an intriguing conjecture:
i$BF$ theories may be interpreted as describing a family of entangled parallel
universes~\cite{wolf1990parallel}. In this picture, intralayer loop excitations
play the role of cosmic strings: they carry invisible charges and do not exert
dynamical forces on particles, yet they generate nontrivial Aharonov--Bohm
phases when enclosed by charged-particle
trajectories~\cite{PhysRevLett.62.1071}. In contrast, interlayer loop excitations
carrying gauge charges can be viewed as wormholes~\cite{PhysRevLett.61.1446}
connecting adjacent universes, as they represent topological defects that alter
the connectivity along the stacking direction. The interlayer $BF$ couplings
then correspond to condensates of such wormholes, with the coupling strength
controlling the effective tunneling amplitude between neighboring universes.
This ``wormhole condensation'' picture is analogous to the anyon-condensation
mechanism in multilayer fractional quantum Hall
systems~\cite{wen2004quantum}. From this perspective, i$BF$ theories with
Toeplitz braiding describe parallel universes whose mutual tunneling is strong
enough that cosmic strings in one universe can imprint robust topological phase
factors on charged particles in another, even when the universes are far apart.

Several directions merit future exploration. A natural next step is to
construct explicit lattice realizations of i$BF$ theories that support Toeplitz
braiding, enabling direct numerical studies and potentially guiding
experimental implementations. Conversely, it is equally worthwhile to develop
field-theoretical approaches, including i$BF$ theory, to revisit topics
extensively studied in lattice settings—such as generalized entanglement
renormalization~\cite{fracton52} and foliation
structures~\cite{fracton16,fracton27,ma2022fractonic}—in the context of
four-dimensional fracton topological orders. In this work we focused on
particle--loop braiding arising from stacking \textit{pure} $BF$ theories.
Extending the construction by incorporating twisted terms to realize multiloop
and Borromean-rings braiding remains an open
problem~\cite{wang_levin1,PhysRevLett.121.061601,2016arXiv161209298P,zhang_compatible_2021,Zhang2023Continuum,Zhang2023fusion,Huang2023,2024arXiv240519077H,huang2025double,zhang2026gsym}.
Another intriguing direction is to incorporate global or subsystem symmetries
into i$BF$ theories and to explore the resulting symmetry-fractionalization
patterns.

\acknowledgments
P.Y. thanks Liujun Zou and Ching Hua Lee for insightful discussions during his
visit to the National University of Singapore. This work was partially supported
by the National Natural Science Foundation of China (NSFC) under Grant
No.~12474149, the Research Center for Magnetoelectric Physics of Guangdong
Province under Grant No.~2024B0303390001, and the Guangdong Provincial Key
Laboratory of Magnetoelectric Physics and Devices under Grant
No.~2022B1212010008.

%

 \appendix
 \section{Detailed calculation on  ZSMs}\label{appendix_detailed}
\subsection{Detailed calculation on ZSMs of HN-type $K$ matrices\label{detailedhn}}
Left singular modes and right singular modes denoted by
$\boldsymbol u$ and $\boldsymbol v$ are determined by
$K_{\mathrm{HN}}\boldsymbol v=\sigma \boldsymbol u$, $K_{\mathrm{HN}}^\mT \boldsymbol u=\sigma \boldsymbol v$, respectively. These equations render
\begin{gather}
	K_{\mathrm{HN}} K_{\mathrm{HN}}^\mT \boldsymbol u=\sigma^2 \boldsymbol u,\quad K_{\mathrm{HN}}^\mT K_{\mathrm{HN}} \boldsymbol v=\sigma^2 \boldsymbol v. \label{equationsvd}
\end{gather}
Therefore, determining the singular modes \( \boldsymbol{u} \) and \(\boldsymbol{v}\) is equivalent to finding the eigenmodes of the Hermitian matrices \(K_{\mathrm{HN}}  K_{\mathrm{HN}} ^\mT\) and \(K_{\mathrm{HN}} ^\mT K_{\mathrm{HN}} \).  One can then adopt the customary technique from the study of boundary zero modes of Hermitian systems.
Since we focus on the ZSMs denoted by $\mathbf u_1$ and $\mathbf v_1$ with exponentially small singular values ($\det K_{\mathrm{HN}}\neq 0$), the
$\sigma\rightarrow 0$ is taken before further calculation. Furthermore, multiplying the equations in Eq.~(\ref{equationsvd}) by $K_{\mathrm{HN}}^{-1}$ and $(K_{\mathrm{HN}}^{-1})^\mT$, respectively, we obtain 
\begin{gather}
	K_{\mathrm{HN}}^\mT\mathbf u_1=0,\quad K_{\mathrm{HN}} \mathbf v_1=0. \label{simplifiedsvdzero}
\end{gather}
Assume the solution of boundary zero mode takes the form
\begin{gather}
    \mathbf u_1 =(u_1,u_2,\ldots,u_N)^\mT,\quad
    \mathbf v_1 =(v_1,v_2,\ldots,v_N)^\mT.
\end{gather}
Eq.~(\ref{simplifiedsvdzero}) renders
\begin{subequations}
\begin{numcases}{}
b u_{j}+n u_{j-1} =0,\quad j=2,3,\ldots,N;\label{bozsvrecurrence4}\\
n u_N =0; \label{bozsvrecurrence2}
\end{numcases}
\end{subequations}
\vspace{-1.3em} 
\begin{subequations}
\begin{numcases}{}
n v_1=0;\label{bozsvrecurrence3}\\
b v_{j-1}+n v_j = 0,\quad j=2,3,\ldots,N. \label{bozsvrecurrence1}
\end{numcases}
\end{subequations}
Eq.~(\ref{bozsvrecurrence4}) is solved by $u_j=\left(-\frac{n}{b}\right)^{j-1}$ for $j=1,\ldots,N$. If $|b|>|n|$, Eq.~(\ref{bozsvrecurrence2}) is only violated by exponentially small terms before taking thermodynamic limit $N\rightarrow\infty$.
  Because eigenspectra and eigenstates of Hermitian matrices are stable under small perturbations of matrix elements, one may slightly adjust the matrix elements to absorb this exponentially small violation. This operation does not severely alter the eigenspectrum or the eigenstates of the original matrix $K_{\mathrm{HN}} K_{\mathrm{HN}}^\mT$. 
  In the thermodynamic limit $N\rightarrow\infty$, Eq.~(\ref{bozsvrecurrence2}) is strictly satisfied.
Therefore, in the case $|b|>|n|$, 
the following normalized solution to LZSM appears:
\begin{gather}
	\mathbf u_1= \sqrt{\frac{1-(\frac{n}{b})^2}{1-(\frac{n}{b})^{2N}}}\begin{pmatrix}
		1 & -\frac{n}{b} & \cdots & \left(-\frac{n}{b}\right)^{N-1}
	\end{pmatrix}^\mT.
\end{gather}
Likewise, in the case $|b|>|n|$, the following normalized solution to RZSM appears:
\begin{gather}
	\mathbf v_1= \sqrt{\frac{1-(\frac{n}{b})^2}{1-(\frac{n}{b})^{2N}}}\begin{pmatrix}
		\left(-\frac{n}{b}\right)^{N-1} & \left(-\frac{n}{b}\right)^{N-2} & \cdots & 1
	\end{pmatrix}^\mT.
\end{gather}

Another point worth noting is that Eq.~(\ref{simplifiedsvdzero}) resembles those used to solve for boundary zero modes of Hermitian matrices. However, the zero modes obtained from Eq.~(\ref{simplifiedsvdzero}) are not necessarily the true zero eigenmodes of $K_{\mathrm{HN}}^\mT$ and $K_{\mathrm{HN}}$, as 
 the eigenspectra and eigenstates of non-Hermitian matrices are highly sensitive  under even exponentially small perturbations of matrix elements.
Therefore, the exponentially small violations may not be absorbed by a slight modification of the matrix elements of $K_{\mathrm{HN}}^\mT$ and $K_{\mathrm{HN}}$.

 \subsection{Detailed calculation on ZSMs of nSSH-type $K$ matrices\label{detailednssh}}
 
Left singular modes and right singular modes denoted by
$\boldsymbol u$ and $\boldsymbol v$ are determined by
$K_{\mathrm{nSSH}}\boldsymbol v=\sigma \boldsymbol u$, $K_{\mathrm{nSSH}}^\mT \boldsymbol u=\sigma \boldsymbol v$, respectively. These equations render
\begin{gather}
	K_{\mathrm{nSSH}} K_{\mathrm{nSSH}}^\mT\boldsymbol u=\sigma^2\boldsymbol u,\quad K_{\mathrm{nSSH}}^\mT K_{\mathrm{nSSH}} \boldsymbol v=\sigma^2 \boldsymbol v. \label{equationsvd2}
\end{gather}
In the thermodynamic limit $N\rightarrow \infty$,  the singular values of ZSMs approach zero, i.e. $\sigma\rightarrow 0$. Since we focus on the ZSMs denoted by $\mathbf u$ and $\mathbf v$ with exponentially small singular values ($\det K_{\mathrm{nSSH}}\neq 0$), the
$\sigma\rightarrow 0$ is taken before further calculation. For the same reason described in Appendix \ref{detailedhn}, the above two equations yield
\begin{gather}
	K_{\mathrm{nSSH}}^\mT\mathbf u=0,\quad K_{\mathrm{nSSH}} \mathbf v=0. \label{simplifiedzsm2}
\end{gather}
Denote the components of $\mathbf v$ and $\mathbf u$ as
\begin{gather}
	 \mathbf u = \begin{pmatrix}
		u_{1,A} & u_{1,B} & \cdots & u_{N,A} & u_{N,B}
	\end{pmatrix}^\mT, \notag \\  \mathbf v = \begin{pmatrix}
		v_{1,A} & v_{1,B} & \cdots & v_{N,A} & v_{N,B}
	\end{pmatrix}^\mT.
\end{gather}
The bulk recurrence relation encoded in the equation $K_{\mathrm{nSSH}}^\mT  \mathbf u=0$ is
\begin{gather}
	\begin{pmatrix}
		0 & m_2 \\ 0 & 0
	\end{pmatrix}\begin{pmatrix}
		u_{j-1,A} \\ u_{j-1,B}
	\end{pmatrix}+\begin{pmatrix}
		0 & n_2 \\ n_1 & 0
	\end{pmatrix}\begin{pmatrix}
		u_{j,A} \\ u_{j,B}
	\end{pmatrix}\notag \\
	+\begin{pmatrix}
		0 & 0 \\ m_1 & 0
	\end{pmatrix}\begin{pmatrix}
		u_{j+1,A} \\ u_{j+1,B}
	\end{pmatrix}=0, \quad j=2,3,\ldots,N-1.
\end{gather}
The ansatz for $\begin{pmatrix}
	u_{j,A} & u_{j,B}
\end{pmatrix}^\mT$ is 
\begin{gather}
	\begin{pmatrix}
	u_{j,A} & u_{j,B}
\end{pmatrix}^\mT=\beta^j \begin{pmatrix}
	u_{A} & u_{B}
\end{pmatrix}^\mT,
\end{gather}
which renders
\begin{gather}
	\begin{pmatrix}
		0 & m_2\beta^{-1}+n_2 \\
		n_1+ m_1\beta & 0
	\end{pmatrix}\begin{pmatrix}
	u_{A} \\ u_{B}
\end{pmatrix}=0.
\end{gather}
The condition for nontrivial solution is
\begin{gather}
	\det\begin{pmatrix}
		0 & m_2\beta^{-1}+n_2 \\
		n_1+ m_1\beta & 0
	\end{pmatrix}=0,
\end{gather}
and the solutions are
\begin{gather}
	\beta_1=-\frac{n_1}{m_1},\quad \beta_2=-\frac{m_2}{n_2}.
\end{gather}
Hence the corresponding solution to $u_A$, $u_B$ is
\begin{gather}
	u_{A,1} =1,\quad  u_{B,1} =0;\\
	u_{A,2} =0,\quad  u_{B,2} =1.
\end{gather}
Denote the corresponding candidate boundary zero mode as $\mathbf u_1$ and $\mathbf u_2$.
If $|\beta_1|<1$, $|m_1|>|n_1|$, then 
\begin{equation}
	\mathbf u_1=\sqrt{\frac{1-\frac{n_1^2}{m_1^2}}{1-(\frac{n_1}{m_1})^{2N}}}\begin{pmatrix}
		 1 & 0 & -\frac{n_1}{m_1} & 0 &\cdots & \left(-\frac{n_1}{m_1}\right)^{N-1} & 0
	\end{pmatrix}^\mT
\end{equation}
is automatically a legitimate solution of LZSM, since the $w_1$ boundary equation is satisfied:
\begin{gather}
	 \begin{pmatrix}
		0 & n_2 \\ n_1 & 0
	\end{pmatrix}\begin{pmatrix} 1 \\0 \end{pmatrix}+\begin{pmatrix}
		0 & 0 \\ m_1 & 0
	\end{pmatrix}\begin{pmatrix} -\frac{n_1}{m_1} \\0 \end{pmatrix}
	=0.
\end{gather}

If $|\beta_2|>1$, $|m_2|>|n_2|$, then
\begin{equation}
	\mathbf u_2=\sqrt{\frac{1-\frac{m_2^2}{n_2^2}}{1-(\frac{m_2}{n_2})^{2N}}}\begin{pmatrix}
		0 & 1 & 0 & -\frac{m_2}{n_2} & \cdots & 0 & \left(-\frac{m_2}{n_2}\right)^{N-1}
	\end{pmatrix}^\mT
\end{equation}
 is automatically a legitimate solution to LZSM, since the $w_2$ boundary equation is satisfied:
 \begin{gather}
 	\begin{pmatrix}
		0 & m_2 \\ 0 & 0
	\end{pmatrix}\left(-\frac{m_2}{n_2}\right)^{N-1}\begin{pmatrix} 0\\1 \end{pmatrix}\notag \\ +\begin{pmatrix}
		0 & n_2 \\ n_1 & 0
	\end{pmatrix}\left(-\frac{m_2}{n_2}\right)^N\begin{pmatrix} 0\\1 \end{pmatrix}=0.
 \end{gather}
 
Hence there are at most two solutions to LZSM: 
\begin{gather}
\mathbf u_1=\sqrt{\frac{1-\frac{n_1^2}{m_1^2}}{1-(\frac{n_1}{m_1})^{2N}}}\begin{pmatrix}
		 1 & 0 & -\frac{n_1}{m_1} & 0 &\cdots & \left(-\frac{n_1}{m_1}\right)^{N-1} & 0
	\end{pmatrix}^\mT, \\
	\mathbf u_2=\sqrt{\frac{1-\frac{m_2^2}{n_2^2}}{1-(\frac{m_2}{n_2})^{2N}}}\begin{pmatrix}
		0 & 1 & 0 & -\frac{m_2}{n_2} & \cdots & 0 & \left(-\frac{m_2}{n_2}\right)^{N-1}
	\end{pmatrix}^\mT.
\end{gather}
If $|m_1|>|n_1|$, then $\mathbf u_1$ is a legitimate solution to LZSM. If $|m_2|>|n_2|$, then $\mathbf u_2$ is automatically a legitimate solution to LZSM.

The discussion to $K_{\mathrm{nSSH}}\mathbf v=0$ is similar.
The bulk recurrence relation encoded in the equation $K_{\mathrm{nSSH}} \mathbf v=0$ is
\begin{gather}
	\begin{pmatrix}
		0 & m_1 \\ 0 & 0
	\end{pmatrix}\begin{pmatrix}
		v_{j-1,A} \\ v_{j-1,B}
	\end{pmatrix}+\begin{pmatrix}
		0 & n_1 \\ n_2 & 0
	\end{pmatrix}\begin{pmatrix}
		v_{j,A} \\ v_{j,B}
	\end{pmatrix}\notag \\
	+\begin{pmatrix}
		0 & 0 \\ m_2 & 0
	\end{pmatrix}\begin{pmatrix}
		v_{j+1,A} \\ v_{j+1,B}
	\end{pmatrix}=0, \quad j=2,3,\ldots,N-1.
\end{gather}
The ansatz for $\begin{pmatrix}
	v_{j,A} & v_{j,B}
\end{pmatrix}^\mT$ is 
\begin{gather}
	\begin{pmatrix}
	v_{j,A} & v_{j,B}
\end{pmatrix}^\mT=\beta^j \begin{pmatrix}
	v_{A} & v_{B}
\end{pmatrix}^\mT,
\end{gather}
which renders
\begin{gather}
	\begin{pmatrix}
		0 & m_1\beta^{-1}+n_1 \\
		n_2+ m_2\beta & 0
	\end{pmatrix}\begin{pmatrix}
	v_{A} \\ v_{B}
\end{pmatrix}=0.
\end{gather}
The condition for nontrivial solution is
\begin{gather}
	\det\begin{pmatrix}
		0 & m_1\beta^{-1}+n_1 \\
		n_2+ m_2\beta & 0
	\end{pmatrix}=0,
\end{gather}
and the solutions are
\begin{gather}
	\beta_1=-\frac{m_1}{n_1},\quad \beta_2=-\frac{n_2}{m_2}.
\end{gather}
Hence the corresponding solution to $v_A$, $v_B$ is
\begin{gather}
	v_{A,1} =0,\quad  v_{B,1} =1;\\
	v_{A,2} =1,\quad  v_{B,2} =0.
\end{gather}
Denote the corresponding candidate boundary zero mode as $\mathbf v_1$ and $\mathbf v_2$.
If $|\beta_1|>1$, $|m_1|>|n_1|$, then $\mathbf v_1$ 
\begin{gather}
	\mathbf v_1=\sqrt{\frac{1-\frac{m_1^2}{n_1^2}}{1-(\frac{m_1}{n_1})^{2N}}}\begin{pmatrix}
		0 & 1 & 0 & -\frac{m_1}{n_1} & \cdots & 0 & \left(-\frac{m_1}{n_1}\right)^{N-1}
	\end{pmatrix}^\mT
\end{gather}
is automatically a legitimate solution of RZSM, since the $w_2$ boundary equation
 \begin{gather}
 	\begin{pmatrix}
		0 & m_1 \\ 0 & 0
	\end{pmatrix}\left(-\frac{m_1}{n_1}\right)^{N-1}\begin{pmatrix} 0\\1 \end{pmatrix}\notag \\ +\begin{pmatrix}
		0 & n_1 \\ n_2 & 0
	\end{pmatrix}\left(-\frac{m_1}{n_1}\right)^N\begin{pmatrix} 0\\1 \end{pmatrix}=0
 \end{gather}
 is satisfied.
 
If $|\beta_2|<1$, $|m_2|>|n_2|$, then
\begin{equation}
	\mathbf v_2=\sqrt{\frac{1-\frac{n_2^2}{m_2^2}}{1-(\frac{n_2}{m_2})^{2N}}}\begin{pmatrix}
		 1 & 0 & -\frac{n_2}{m_2} & 0 &\cdots & \left(-\frac{n_2}{m_2}\right)^{N-1} & 0
	\end{pmatrix}^\mT
\end{equation}
is automatically a legitimate solution to RZSM, since the $w_1$ boundary equation is automatically satisfied:
\begin{gather}
	 \begin{pmatrix}
		0 & n_1 \\ n_2 & 0
	\end{pmatrix}\begin{pmatrix} 1\\0 \end{pmatrix}+\begin{pmatrix}
		0 & 0 \\ m_2 & 0
	\end{pmatrix}\begin{pmatrix} -\frac{n_2}{m_2}\\0 \end{pmatrix}
	=0.
\end{gather}
Hence there are at most two solutions to RZSMs:
\begin{gather}
	\mathbf v_1=\sqrt{\frac{1-\frac{m_1^2}{n_1^2}}{1-(\frac{m_1}{n_1})^{2N}}}\begin{pmatrix}
		0 & 1 & 0 & -\frac{m_1}{n_1} & \cdots & 0 & \left(-\frac{m_1}{n_1}\right)^{N-1}
	\end{pmatrix}^\mT, \\
	\mathbf v_2=\sqrt{\frac{1-\frac{n_2^2}{m_2^2}}{1-(\frac{n_2}{m_2})^{2N}}}\begin{pmatrix}
		 1 & 0 & -\frac{n_2}{m_2} & 0 &\cdots & \left(-\frac{n_2}{m_2}\right)^{N-1} & 0
	\end{pmatrix}^\mT.
\end{gather}
In the case $|m_1|>|n_1|$, $\mathbf v_1$ is a legitimate RZSM. In the case $|m_2|>|n_2|$, $\mathbf v_2$ is a legitimate RZSM.

\end{document}